\documentclass[conference]{IEEEtran}

\usepackage{tikz}
\usepackage{amsmath}

\usepackage{color}
\usepackage{xcolor}
\usepackage{amssymb}
\usepackage{url}
\usepackage{caption}
\usepackage{subcaption}
\usepackage{float}
\usepackage{balance}
\usepackage{graphicx}
\usepackage{textcomp}
\usepackage{balance}
\usepackage{booktabs} 
\usepackage[pagebackref=true,breaklinks=true,colorlinks,bookmarks=false]{hyperref}

\DeclareMathOperator*{\argmax}{arg\,max}
\newcommand{\user}{u}
\newcommand{\User}{\mathcal{U}}
\newcommand{\Edge}{\mathcal{E}}
\newcommand{\Graph}{\mathcal{G}}
\newcommand{\proxi}{s}
\newcommand{\friend}{\kappa}
\newcommand{\ano}{\mathcal{A}}
\newcommand{\rec}{\mathcal{R}}
\newcommand{\Neighbor}{\omega}
\newcommand{\entropy}{\hat H}

\newcommand{\dk}{\ensuremath{D}}
\newcommand{\dkelement}{\mathit{r}}
\newcommand{\noise}{\zeta}
\newcommand{\noisesample}{\bar{\zeta}}
\newcommand{\Fake}{\mathcal{B}}
\newcommand{\Proxi}{\mathcal{S}}
\newcommand{\weight}{w}

\newcommand{\degdiff}{\Delta}

\newcommand{\candi}{\gamma}
\newcommand{\candiproxi}{\lambda}
\newcommand{\fix}{\mathcal{F}}

\newcommand{\mypara}[1]{\medskip\noindent{\bf {#1}:}~}

\begin{document}

\title{Towards Plausible Graph Anonymization}

\author{
\IEEEauthorblockN{
Yang Zhang\IEEEauthorrefmark{1}, 
Mathias Humbert\IEEEauthorrefmark{2}, 
Bartlomiej Surma\IEEEauthorrefmark{1},
\\
Praveen Manoharan\IEEEauthorrefmark{1},
Jilles Vreeken\IEEEauthorrefmark{1},
Michael Backes\IEEEauthorrefmark{1}
} 
\IEEEauthorblockA{\IEEEauthorrefmark{1}CISPA Helmholtz Center for Information Security,
\\
\{zhang, bartlomiej.surma, praveen.manoharan, vreeken, backes\}@cispa.saarland}
\IEEEauthorblockA{\IEEEauthorrefmark{2}Cyber-Defence Campus, armasuisse Science and Technology,
mathias.humbert@armasuisse.ch}
}

\IEEEoverridecommandlockouts
\makeatletter\def\@IEEEpubidpullup{6.5\baselineskip}\makeatother
\IEEEpubid{\parbox{\columnwidth}{
    Network and Distributed Systems Security (NDSS) Symposium 2020\\
    23-26 February 2020, San Diego, CA, USA\\
    ISBN 1-891562-61-4\\
    https://dx.doi.org/10.14722/ndss.2020.23xxx\\
    www.ndss-symposium.org
}
\hspace{\columnsep}\makebox[\columnwidth]{}}

\maketitle
\begin{abstract}
Social graphs derived from online social interactions contain a wealth of information that is nowadays extensively used by both industry and academia. However, as social graphs contain sensitive information, they need to be properly anonymized before release. Most of the existing graph anonymization mechanisms rely on the perturbation of the original graph's edge set. In this paper, we identify a fundamental weakness of these mechanisms: They neglect the strong structural proximity between friends in social graphs, thus add implausible fake edges for anonymization. 

To exploit this weakness, we first propose a metric to quantify an edge's plausibility by relying on graph embedding. Extensive experiments on three real-life social network datasets demonstrate that our plausibility metric can very effectively differentiate fake edges from original edges with AUC (area under the ROC curve) values above 0.95 in most of the cases. We then rely on a Gaussian mixture model to automatically derive the threshold on the edge plausibility values to determine whether an edge is fake, which enables us to recover to a large extent the original graph from the anonymized graph. We further demonstrate that our graph recovery attack jeopardizes the privacy guarantees provided by the considered graph anonymization mechanisms. 

To mitigate this vulnerability, we propose a method to generate fake yet plausible edges given the graph structure and incorporate it into the existing anonymization mechanisms. Our evaluation demonstrates that the enhanced mechanisms decrease the chances of graph recovery, reduce the success of graph de-anonymization (up to 30\%), and provide even better utility than the existing anonymization mechanisms.
\end{abstract}

\section{Introduction}
\label{sec:intro}

The rapid development of online social networks (OSNs)
has resulted in an unprecedented scale of social graph data available. 
Access to such data is invaluable for both the industrial and academic domains. 
For instance, Amazon or Netflix have leveraged graph data to improve 
their recommendation services.
Moreover, researchers have been using graph data to gain a deeper understanding 
of many fundamental societal questions,
such as people's communication patterns~\cite{OSHSLKKB07,Z19},
geographical movement~\cite{CML11,ZHRLPB18},
and information propagation~\cite{KKT03,RMK11}.
These examples demonstrate that the sharing of large-scale graph data can bring significant benefits to the society.

On the downside, graph data also inherently contains very sensitive information about individuals~\cite{BL18},
such as their social relations~\cite{BHPZ17}, 
and it can be used to infer private attributes~\cite{JWZG17}. 
In order to mitigate privacy risks, 
it is crucial to properly anonymize the graph data before releasing it to third parties. 
The naive approach of replacing real identifiers by random numbers 
has been proven ineffective by Backstrom et al. about a decade ago already~\cite{BDK07}.
From then on, the research community has been working on developing 
more robust graph anonymization mechanisms~\cite{LT08,SZWZZ11,MPS13,XCT14,JMB16}.
The majority of the proposed mechanisms 
focus on perturbing the original edge set of the graph 
(instead of perturbing the node set) by adding \emph{fake edges} between users, such that the \emph{perturbed graph} 
satisfies well-established privacy guarantees,
such as $k$-anonymity~\cite{S02} and differential privacy~\cite{DR14}.

\begin{figure*}[!t]
\centering
\begin{subfigure}{0.95\columnwidth}
\includegraphics[width=\columnwidth]{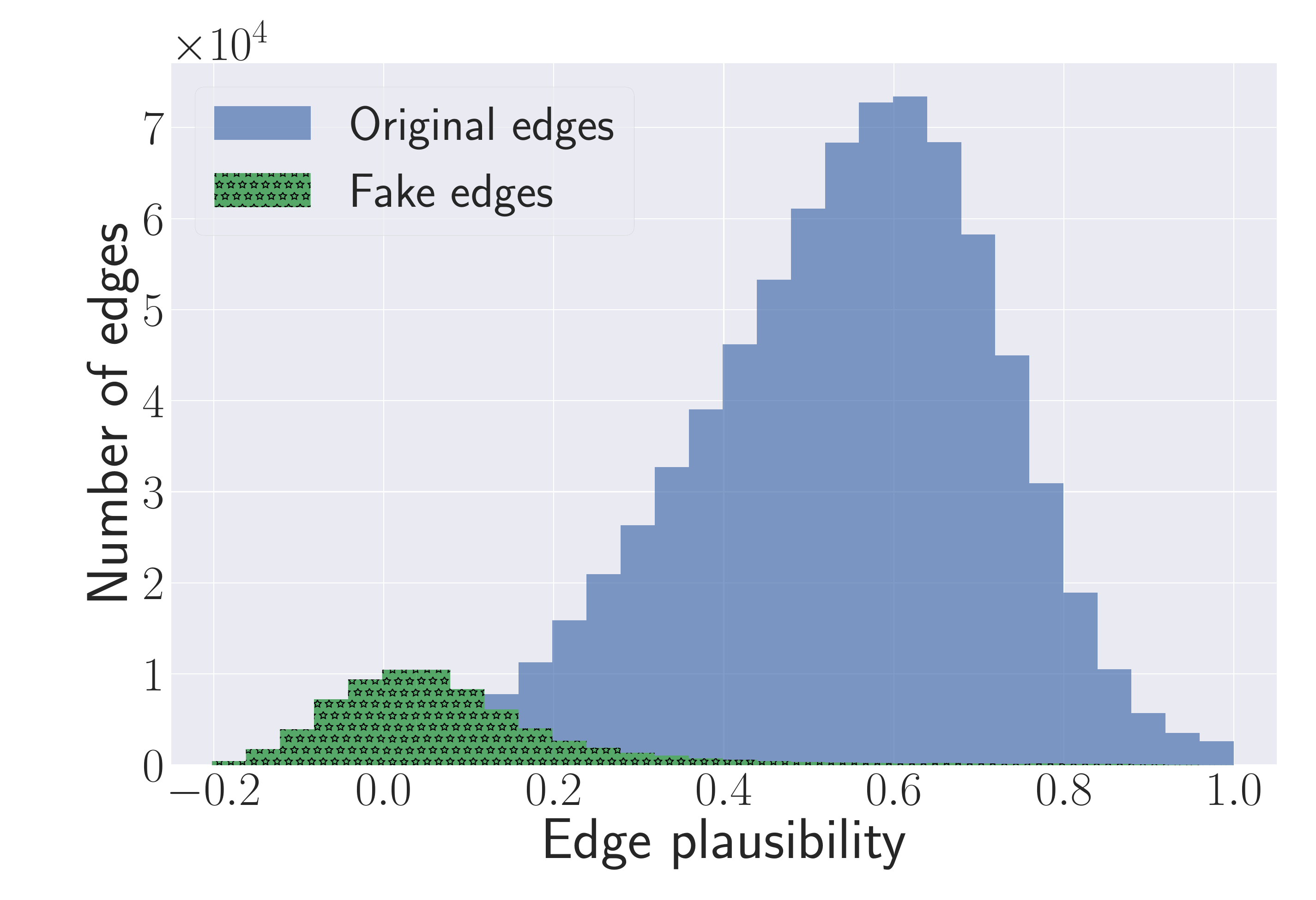}
\caption{$k$-DA ($k=100$)}
\label{fig:facebook_kDa_100_dist_hist}
\end{subfigure}
\begin{subfigure}{0.95\columnwidth}
\includegraphics[width=\columnwidth]{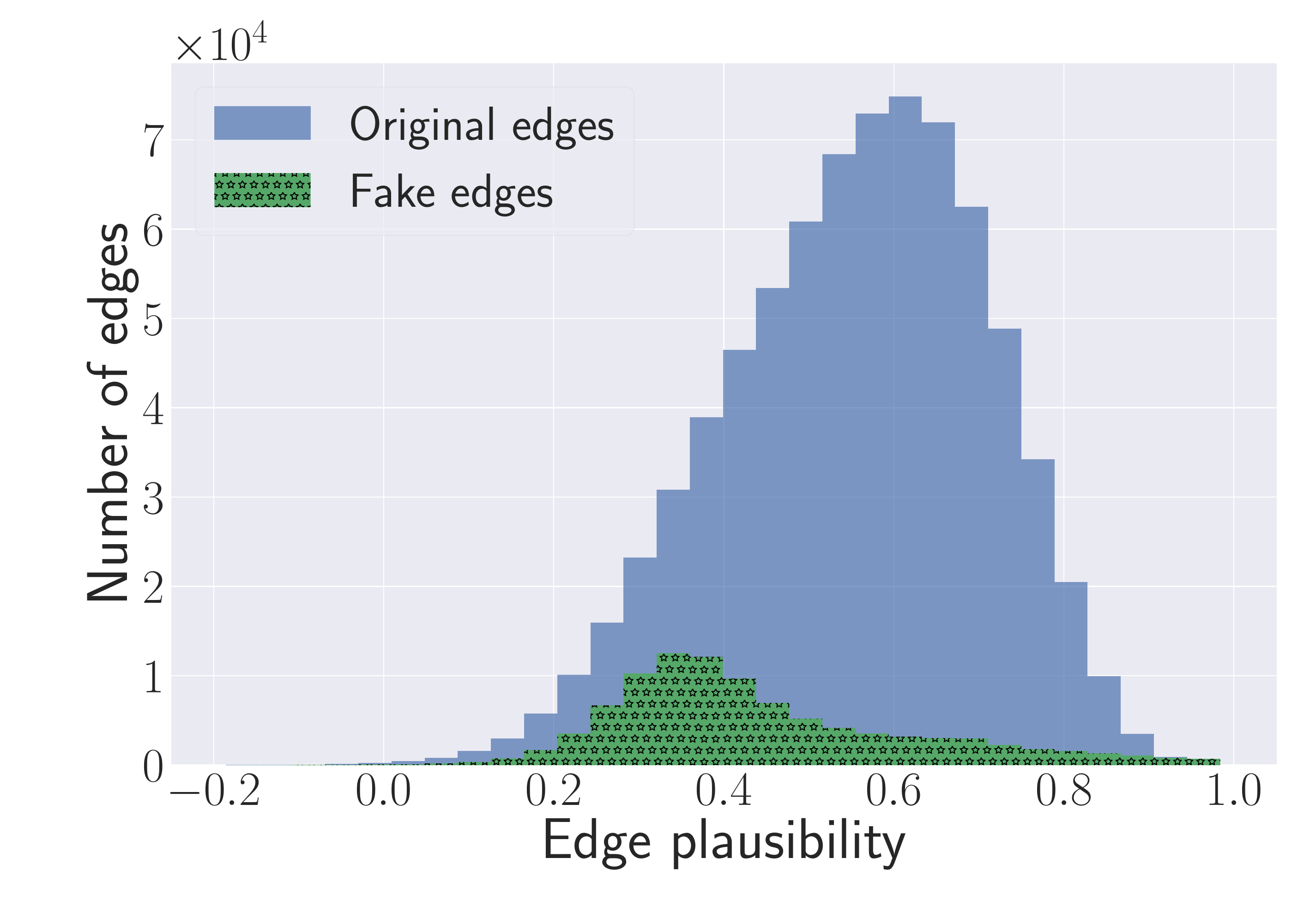}
\caption{Enhanced $k$-DA ($k=100$)}
\label{fig:facebook_ikDa_100_dist_hist}
\end{subfigure}
\caption{Plausibility distributions of fake and original edges 
in the NO dataset anonymized by (a) the original $k$-DA 
and (b) by our enhanced $k$-DA mechanisms.
The edge plausibility is defined in \autoref{sec:proxi}.
The NO dataset is collected by Viswanath et al.~\cite{VMCG09},
and $k$-DA~\cite{LT08} is one of the anonymization mechanisms we concentrate on in this paper.}
\label{fig:dist_hist}
\end{figure*} 

\subsection{Contributions}

In this paper, we identify a fundamental weakness 
of the most prominent graph anonymization mechanisms:
When creating fake edges,
they do not take into account key characteristics of the underlying graph structure,
in particular, the higher structural proximity between friends~\cite{LK07},
which results in fake edges not being plausible enough compared to the original ones.
To exploit this weakness,
we first assess the plausibility of each edge by relying on graph embedding~\cite{PAS14,GL16}.
We show that this approach can very effectively detect fake edges
(see \autoref{fig:facebook_kDa_100_dist_hist} for an example of the edge plausibility distribution of fake and original edges
on a real-life social network dataset),
and thus can eventually help recover the original graph to a large extent.
We then demonstrate that our graph recovery attack 
jeopardizes the anonymization mechanisms' privacy guarantees.
Finally, we develop enhanced versions of the existing graph anonymization mechanisms that: 
(i) create plausible edges (\autoref{fig:facebook_ikDa_100_dist_hist}), 
(ii) reduce the risk of graph recovery and graph de-anonymization,
(iii) preserve the initial privacy criteria provided by the mechanisms, 
and (iv) provide even better graph utility 
(with respect to how well the anonymized graph preserves the structural properties of the original graph).

To illustrate the wide applicability of our approach, 
we concentrate on two of the best established graph anonymization mechanisms,
namely $k$-DA~\cite{LT08} and SalaDP~\cite{SZWZZ11},
which provide $k$-anonymity and differential privacy guarantees, respectively.
The reason we choose $k$-DA and SalaDP is that they are the best graph anonymization schemes 
with respect to utility and resistance to de-anonymization (in addition to being the most cited). 
This conclusion is drawn from the evaluation performed by Ji et al.~\cite{JLMHB15}.

In the following, we provide an overview of our contributions in this paper.

\mypara{Edge Plausibility}
We measure the plausibility of an edge
as the structural proximity between the two users it connects.
In the field of link prediction~\cite{LK07}, 
structural proximity is normally measured by human-designed metrics,
which only capture partial information of the proximity.
Instead, we rely on graph embedding~\cite{PAS14,GL16}
to map users in the anonymized graph into a continuous vector space,
where each user's vector comprehensively reflects her structural properties in the graph.
Then, we define each edge's plausibility as the similarity between the vectors of the two users this edge connects, 
and postulate that lower similarity implies lower edge plausibility.

\mypara{Graph Recovery}
We show the effectiveness of our approach in differentiating fake edges from original ones 
without determining a priori a specific decision threshold on the plausibility metric.
For this case, we adopt the AUC (area under the ROC curve) value 
as the  evaluation metric.
Extensive experiments performed on three real-life social network datasets
show that our plausibility metric achieves excellent performance 
(corresponding to AUC values greater than 0.95) in most of the cases.
Then, observing that the fake and real edges' empirical plausibility follow different Gaussian distributions,
we rely on a Gaussian mixture model and maximum a posteriori probability estimate 
to automatically determine the threshold on the edge plausibility values to detect fake edges.
Our experimental results show that 
this approach achieves strong performance
with F1 scores above 0.8 in multiple cases.
After deleting the fake edges,
we are able to recover, to a large extent, the original graph 
from the anonymized one.

\mypara{Privacy Damage}
The two anonymization mechanisms we consider
follow different threat models and privacy definitions.
To precisely quantify the privacy impact 
of our graph recovery,
we propose privacy loss measures tailored to each mechanism we target.
As the first anonymization mechanism assumes
the adversary uses the users' degrees to conduct her attack,
we evaluate the corresponding privacy impact as the difference 
between users' degrees in the original, anonymized, and recovered graphs. 
For the differential privacy mechanism,
we measure the magnitude and entropy of noise 
added to the statistical measurements of the graph. 
Our experimental results show that the privacy provided 
by both mechanisms significantly decreases,
which demonstrates the vulnerabilities of 
existing graph anonymization techniques.

\mypara{Enhancing Graph Anonymization}
In order to improve the privacy situation, 
we propose a method that generates plausible edges 
while preserving the original privacy guarantees of each mechanism.
We rely on statistical sampling to select potential fake edges
that follow a similar plausibility distribution as the edges in the original graph. 
Our experimental results show that
our enhanced anonymization mechanisms
are less prone to graph recovery (AUC dropping by up to 35\%)
and preserve higher graph utility 
compared to the existing anonymization mechanisms.
More importantly, we show that our enhanced mechanisms reduce
the state-of-the-art graph de-anonymization~\cite{NS09} attack's performance significantly 
(up to 30\% decrease in the number of de-anonymized users).

\medskip
In summary, we make the following contributions in this paper:
\begin{itemize}
\item We perform a graph recovery attack on anonymized social graphs based on graph embedding 
that captures the structural proximity between users and thus unveils fake edges (i.e., relations) between them. 
\item We show through extensive experimental evaluation 
that our graph recovery attack jeopardizes the privacy guarantees
provided in two prominent graph anonymization mechanisms.
\item We propose enhanced versions of these graph anonymization mechanisms that improve both their privacy and utility provisions.
\end{itemize}

\subsection{Organization}
The rest of the paper is organized as follows.
We introduce the notations, anonymization mechanisms,
and threat model used throughout the paper in \autoref{sec:preli}.
\autoref{sec:proxi} presents our edge plausibility definition
and \autoref{sec:evalua} evaluates its effectiveness.
The privacy impact of our graph recovery is studied in \autoref{sec:prideg}.
In \autoref{sec:fix}, we introduce our enhanced graph anonymization mechanisms.
\autoref{sec:relwork} discusses the related work in the field
and \autoref{sec:conclu} concludes the paper.
\section{Preliminaries}
\label{sec:preli}

In this section, we first introduce the notations,
second, describe the two anonymization mechanisms we study, 
and third, present the threat model.

\begin{figure*}[!t]
\centering
\includegraphics[width=2\columnwidth]{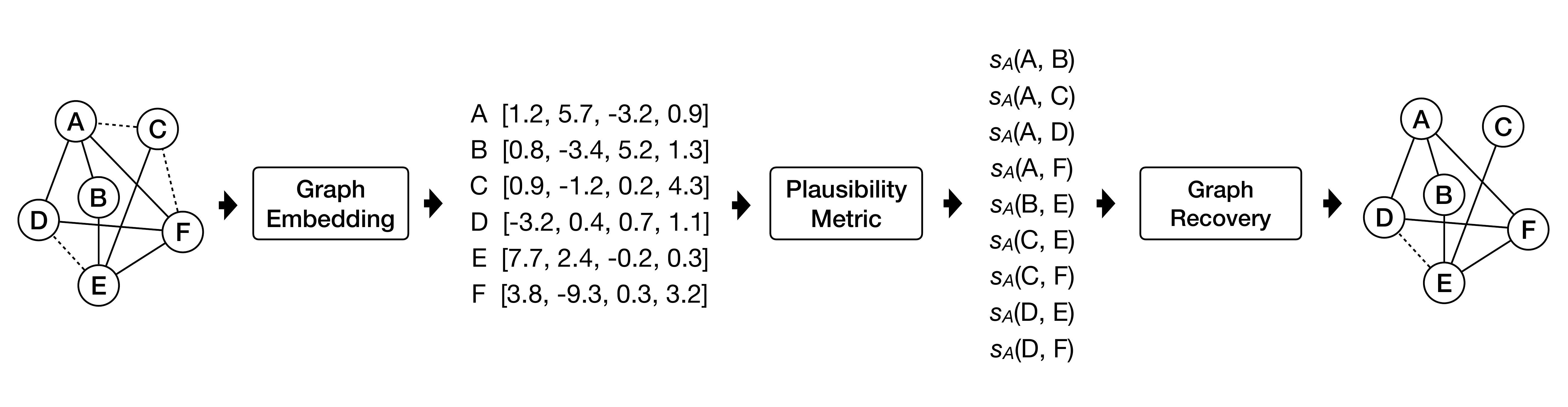}
\caption{A schematic view of our graph recovery attack.
Dashed lines in the graph represent fake added edges;
$\proxi_{\ano}(\text{A}, \text{B})$ represents the plausibility of edge $\{\text{A}, \text{B}\}$ (\autoref{sec:proxi});
A [1.2, 5.7, -3.2, 0.9] represents an example embedding vector of A (\autoref{sec:proxi}).}
\label{fig:system}
\end{figure*} 

\subsection{Notations}

We model a social graph as an undirected graph 
$\Graph=(\User, \Edge)$, where set $\User$ contains the users (nodes) and 
set $\Edge\subseteq \{ \{ \user, \user' \} \vert \user, \user' \in \User \wedge \user\neq\user' \}$ 
represents all the edges of the graph.
We define by $\ano$ the anonymization mechanism
which transforms $\Graph$ to an anonymized graph 
$\Graph_{\ano} = (\User, \Edge_{\ano})$ following the privacy criteria of $\ano$.
By this definition, we only consider graph anonymization mechanisms
that do not add new nodes but only modify edges.
This is in line with most of the previous works~\cite{LT08,ZP08,SZWZZ11,MPS13,XCT14}.
We further use $\friend(\user)$ to represent $\user$'s friends in $\Graph$,
i.e., $\friend(\user) = \{ \user' \vert \{\user, \user'\} \in \Edge \}$.
Accordingly, $\friend_{\ano}(\user)$ represents $\user$'s friends in $\Graph_{\ano}$.
For presentation purposes,
we summarize the notations introduced here and in the following sections in \autoref{table:notations}.

\begin{table}[!t]
\centering
\caption{Notations.}
\label{table:notations}
\begin{tabular}{l | l}
\toprule
Notation & Description\\
\midrule
$\Graph=(\User, \Edge)$ & Social graph\\
$\user\in\User$ & A social network user\\
$\{\user, \user'\}\in \Edge$ & An edge connecting users $\user$ and $\user'$\\
$\ano$ & Anonymization mechanism\\
$\Graph_{\ano}$ & Anonymized social graph\\
$\friend(\user)$ & Friends of user $\user$\\
$f(\user)$ & Embedding vector of user $\user$\\
$\proxi_{\ano}(\user, \user')$ & Plausibility of edge $\{\user, \user'\}$ in $\Graph_{\ano}$\\
$\Graph_{\rec}$ & Recovered social graph\\
$\dk(\Graph)$ & $dK$-2 series of $\Graph$\\
$\Graph_{\fix}$ & Anonymized graph by enhanced mechanism\\
\bottomrule
\end{tabular}
\end{table}

\subsection{Graph Anonymization Mechanisms}

We briefly review the two graph anonymization mechanisms we study.
For more details, we refer the readers to the original papers.
Note that, to fully understand these two mechanisms,
we have also inspected the source code of SecGraph~\cite{JLMHB15},
a state-of-the-art software system for evaluating graph anonymization
which includes an implementation of both $k$-DA and SalaDP.

\mypara{$k$-DA~\cite{LT08}}
\emph{k}-DA follows the notion of $k$-anonymity in database privacy.
The mechanism assumes that the adversary has prior knowledge 
of its target users' degrees in a social graph, i.e., numbers of friends,
and uses this knowledge to identify the targets from the graph.
To mitigate this privacy risk,
\emph{k}-DA modifies the original social graph, 
such that in the resulting anonymized graph,
each user shares the same degree with at least $k-1$ other users.

$k$-DA takes two steps to achieve its goal.
First, it utilizes dynamic programming
to construct a $k$-anonymous degree sequence.
Second, the mechanism adds edges\footnote{In its relaxed version, 
$k$-DA also deletes a small fraction of edges, but its
major operation is still adding edges.} 
to the original graph in order to realize the $k$-anonymous degree sequence. 
By calculating the differences between the original degree sequence 
and the $k$-anonymous degree sequence,
$k$-DA maintains a list that stores the number of edges needed for each user, 
namely the user's \emph{residual degree}.
When adding an edge for a certain user,
$k$-DA picks the new adjacent user with the highest residual degree.

\mypara{SalaDP~\cite{SZWZZ11}}
SalaDP is one of the first and most widely known mechanisms
applying differential privacy for graph anonymization. 
The statistical metric SalaDP concentrates on 
is the $dK$-2 series of a graph $\Graph$ 
which counts, for each pair $(i,j)$ of node degrees $i$ and $j$, 
the number of edges in $\Graph$ that connect nodes of these degrees.
A formal definition of $dK$-2 series will be provided in \autoref{sec:prideg}.

SalaDP also takes a two-step approach to anonymize a graph.
First, the mechanism adds Laplace noise
to each element in the original $dK$-2 series,
and obtains a differentially private $dK$-2 series.
Then, it generates the anonymized graph following the new $dK$-2 series.
By checking SecGraph's source code, 
we find that SalaDP generates the anonymized graph 
by (mainly) adding fake edges
to the original graph
in a random manner.\footnote{Line 252 
of \path{SalaDP.java} in \path{src/anonymize/} of SecGraph.}

\medskip
From the above descriptions,
we can see that neither of the anonymization mechanisms
consider friends' strong structural proximity when adding fake edges.
The main hypothesis we investigate
is that we can effectively differentiate the fake edges added 
by such mechanisms from the original edges, 
using a suitable measure for edge plausibility.
We focus on fake added edges (and not on deleted edges)
since most of the graph anonymization mechanisms
mainly add edges to the original social graph for preserving better graph utility.
It is worth noting that our approach (\autoref{sec:proxi})
can also help recover deleted edges on anonymized graphs.
However, the underlying search space is then $\mathcal{O}(\vert\User \vert^2)$,
which is computationally very expensive on large graphs.
In the future, we plan to tackle this problem by designing heuristics to efficiently recover the deleted edges.

\subsection{Threat Model}

The adversary's goal is to detect fake edges in $\Graph_{\ano}$,
partially recover the original graph,
and eventually carry out privacy attacks on the recovered graph.
To perform graph recovery, 
we assume that the adversary only 
has access to the anonymized graph $\Graph_{\ano}$
and is aware of the underlying anonymization algorithm.
This means that the adversary does not need any information 
about the original graph $\Graph$, 
such as $\Graph$'s graph structure
or any statistical background knowledge related to this graph.
\autoref{fig:system} depicts a schematic overview of the attack.

Our graph recovery attack could also 
be carried out by a service provider (e.g., OSN operator) 
to check whether there are potential flaws 
in its anonymized graph data before release.
\section{Edge Plausibility}
\label{sec:proxi}

To verify our hypothesis that an edge is fake 
if the users it connects are structurally distant,
we first need to quantify two users' structural proximity in a social graph.
Previous work on link prediction provides
numerous proximity metrics~\cite{LK07,BL11,AB17}.
However, these metrics are manually designed
and only capture partial information of structural proximity.
The recent advancement of graph embedding
provides us with an alternative approach~\cite{PAS14,TQWZYM15,GL16,BHPZ17,HYL17,RSBZ19}.
In this context, users in a social network 
are embedded into a continuous vector space,
such that each user's vector 
comprehensively reflects her structural property in the network.
Then, for an edge in the anonymized graph,
we can define its two users' structural proximity as the similarity of their vectors,
and use this similarity as the edge's plausibility.

In this section, we first recall the methodology of graph embedding,
and then formally define edge plausibility.

\subsection{Graph Embedding}

Graph embedding aims to learn a map $f$
from users in $\Graph_{\ano}$ 
to a continuous vector space, i.e.,
\[
f: \User \rightarrow\mathbb{R}^d,
\]
where $d$, as a hyperparameter,
is the dimension of each user's vector.
We adopt the state-of-the-art optimization framework,
namely Skip-gram~\cite{MCCD13,MSCCD13},
to learn $f$; the corresponding objective function is defined as:
\begin{equation}
\label{equ:skipgram}
\argmax_f \prod_{\user \in \User}\prod_{\user' \in \Neighbor(\user)}{P(\user' \vert f(\user))}.
\end{equation}
Here, the conditional probability $P(\user'\vert f(\user))$
is modeled with a softmax function
\[
P(\user'\vert f(\user)) = \frac{\exp(f(\user')\cdot f(\user))}{\sum_{v \in \User}\exp(f(v)\cdot f(\user))},
\]
where $f(\user')\cdot f(\user)$ is the dot product of the two vectors,
and $\Neighbor(\user)$ represents $\user$'s \emph{neighborhood} in $\Graph_{\ano}$.
To define $\Neighbor(\user)$,
we use a random walk approach following previous works~\cite{PAS14, GL16}.
Concretely, we start a random walk from each user in $\Graph_{\ano}$
for a fixed number of times $t$, referred to as the \emph{walk times}.
Each random walk takes $l$ steps, referred to as the \emph{walk length}.
The procedure results in a set of truncated random walk traces,
and each user's neighborhood
includes the users that appear before and after her\footnote{We select 
10 users before and after the considered user following previous works~\cite{PAS14,GL16,BHPZ17}.}
in all these random walk traces.
Similar to the vector dimension ($d$),
walk length and walk times ($l$ and $t$) are also hyperparameters.
We will choose their values experimentally.

Objective function~(\autoref{equ:skipgram}) implies that
if two users share similar neighborhoods in $\Graph_{\ano}$,
then their learned vectors will be closer
than those with different neighborhoods.
This results in each user's vector being able to preserve her neighborhood
and to eventually reflect her structural property in $\Graph_{\ano}$.
To optimize \autoref{equ:skipgram},
we rely on stochastic gradient descent (SGD)
with negative sampling~\cite{MSCCD13}.
We omit the details here due to space limitation.

\subsection{Quantifying Edge Plausibility}

Given the vectors learned from graph embedding,
we define an edge plausibility as the cosine similarity between its two users' vectors.
Formally, for $\{\user, \user'\}\in \Edge_{\ano}$,
its plausibility is defined as:
\[
\proxi_{\ano}(\user, \user') = \frac{f(\user)\cdot f(\user')}{\vert\vert f(\user) \vert\vert_2\ \vert\vert f(\user') \vert\vert_2}
\]
where $\vert\vert \cdot \vert\vert_2$ denotes the $L_2$-norm.
Consequently, if the vectors of two users have higher (cosine) similarity,
then the edge connecting these users is more plausible.
It is worth noting that as $f(\user)\in \mathbb{R}^d$, 
the range of $\proxi_{\ano}(\user, \user')$ lies in [-1, 1] instead of [0, 1].

\begin{figure*}[!ht]
\centering
\begin{subfigure}{0.67\columnwidth}
\includegraphics[width=\columnwidth]{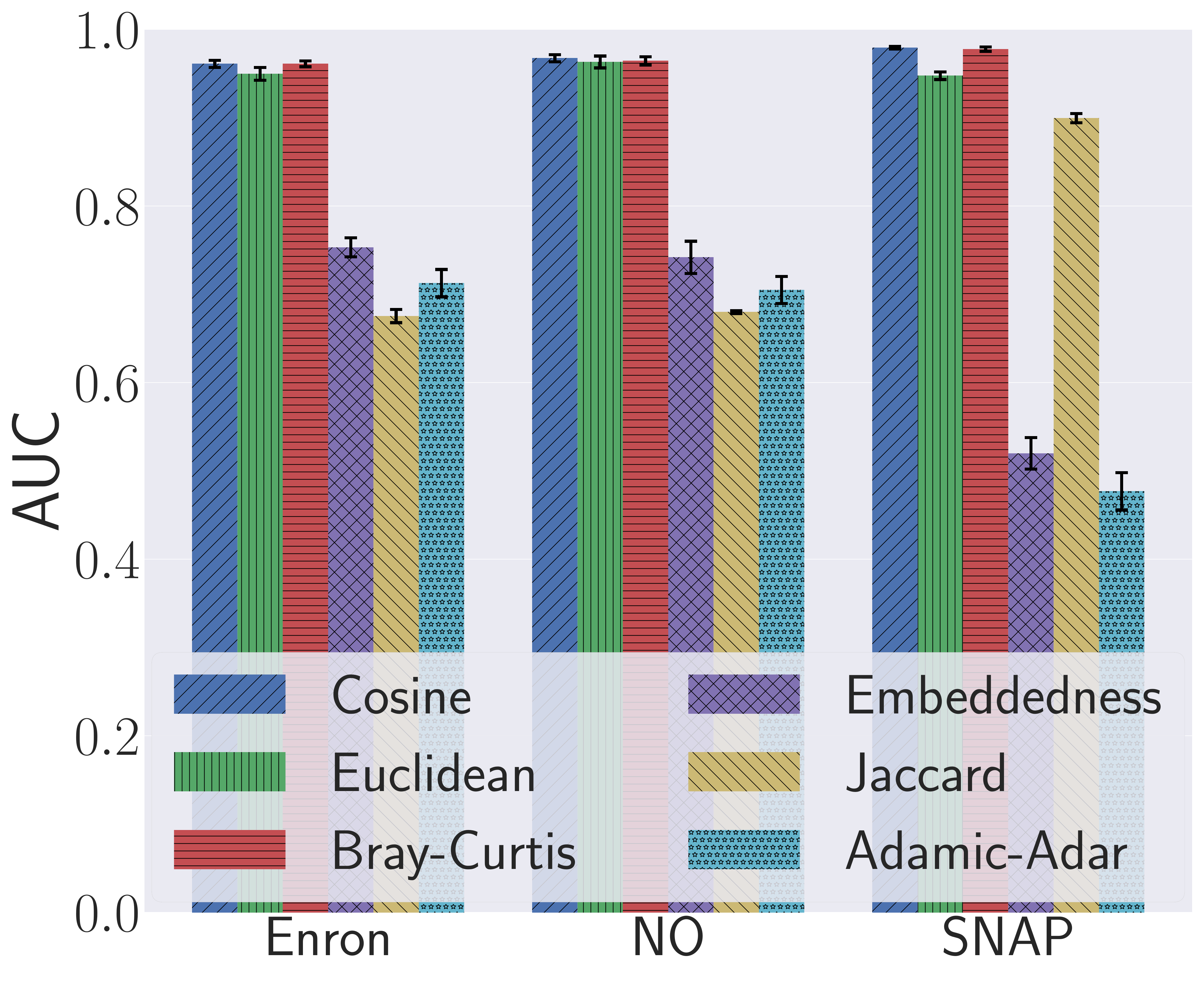}
\caption{$k$-DA ($k=50$)}
\label{fig:kDa_50_auc}
\end{subfigure}
\begin{subfigure}{0.67\columnwidth}
\includegraphics[width=\columnwidth]{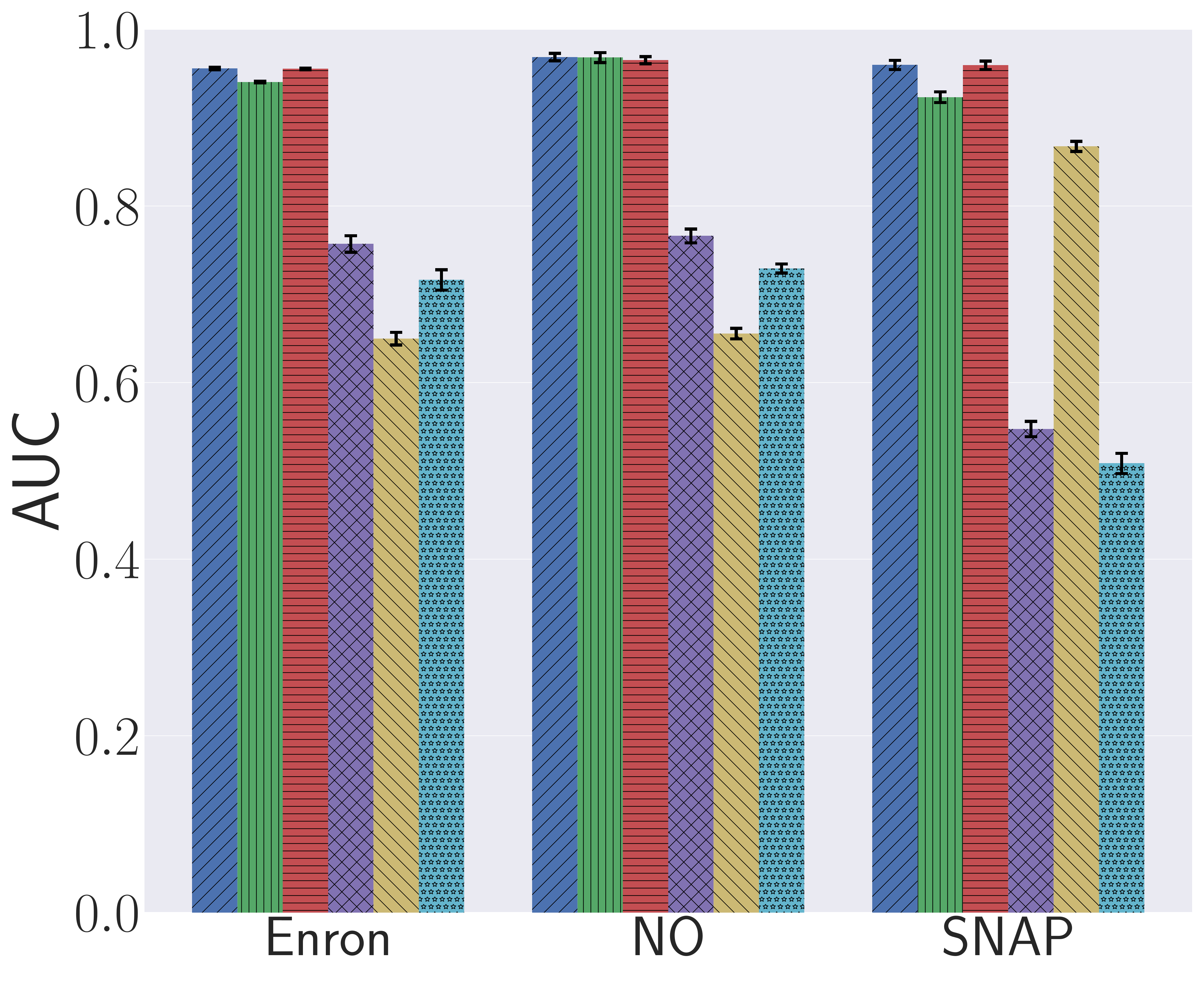}
\caption{$k$-DA ($k=75$)}
\label{fig:kDa_75_auc}
\end{subfigure}
\begin{subfigure}{0.67\columnwidth}
\includegraphics[width=\columnwidth]{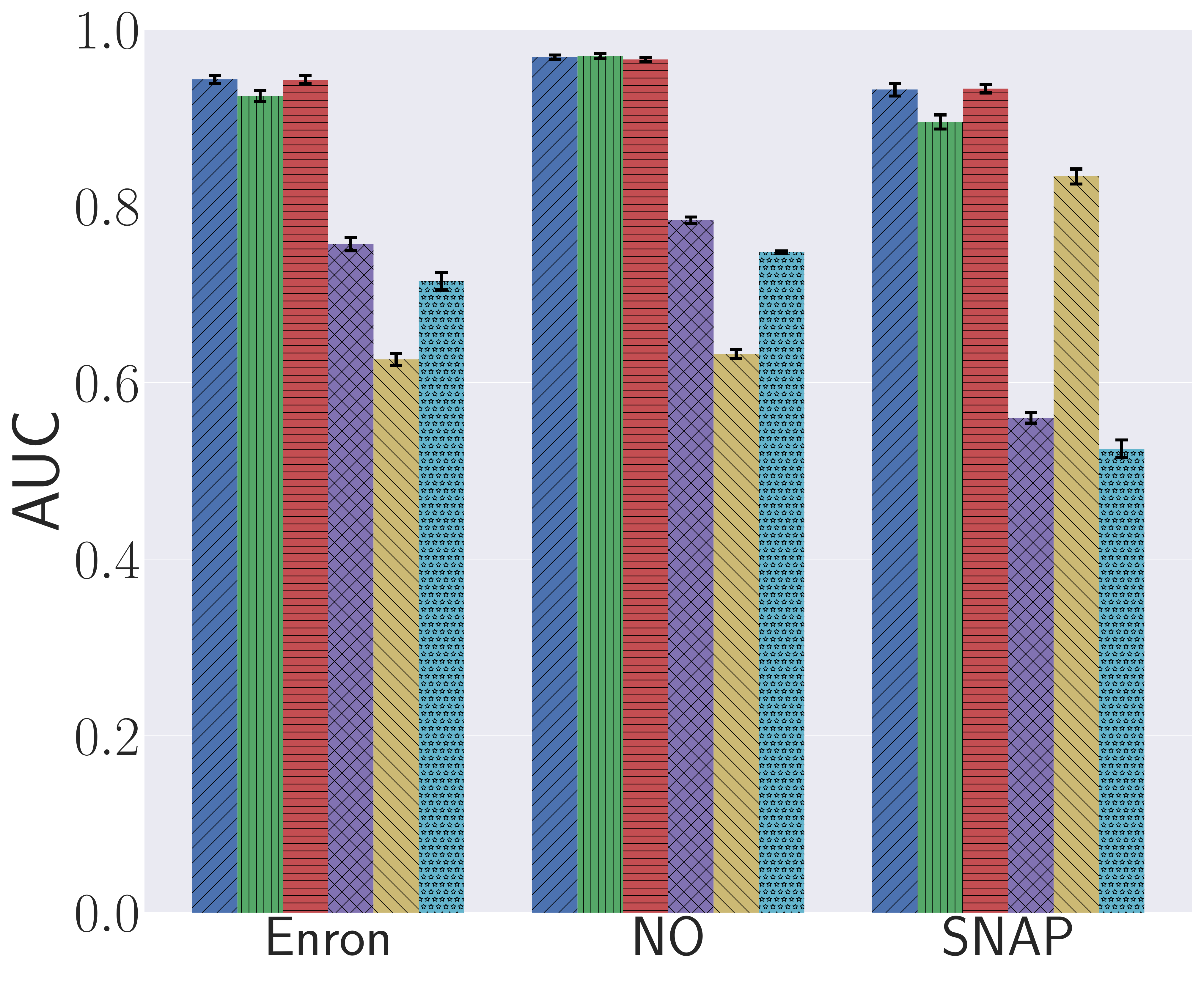}
\caption{$k$-DA ($k=100$)}
\label{fig:kDa_100_auc}
\end{subfigure}
\begin{subfigure}{0.67\columnwidth}
\includegraphics[width=\columnwidth]{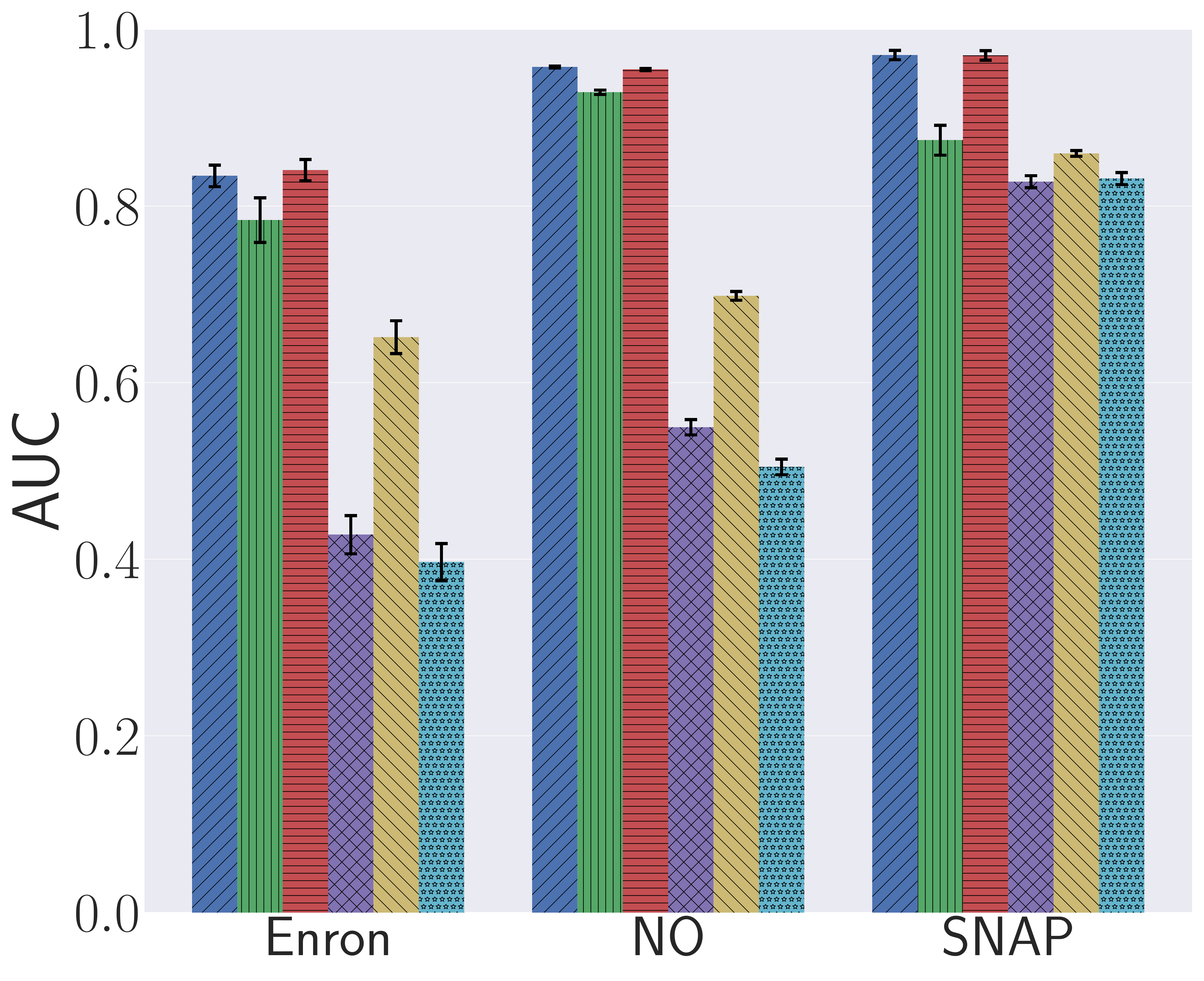}
\caption{SalaDP ($\epsilon=100$)}
\label{fig:sdp_100_auc}
\end{subfigure}
\begin{subfigure}{0.67\columnwidth}
\includegraphics[width=\columnwidth]{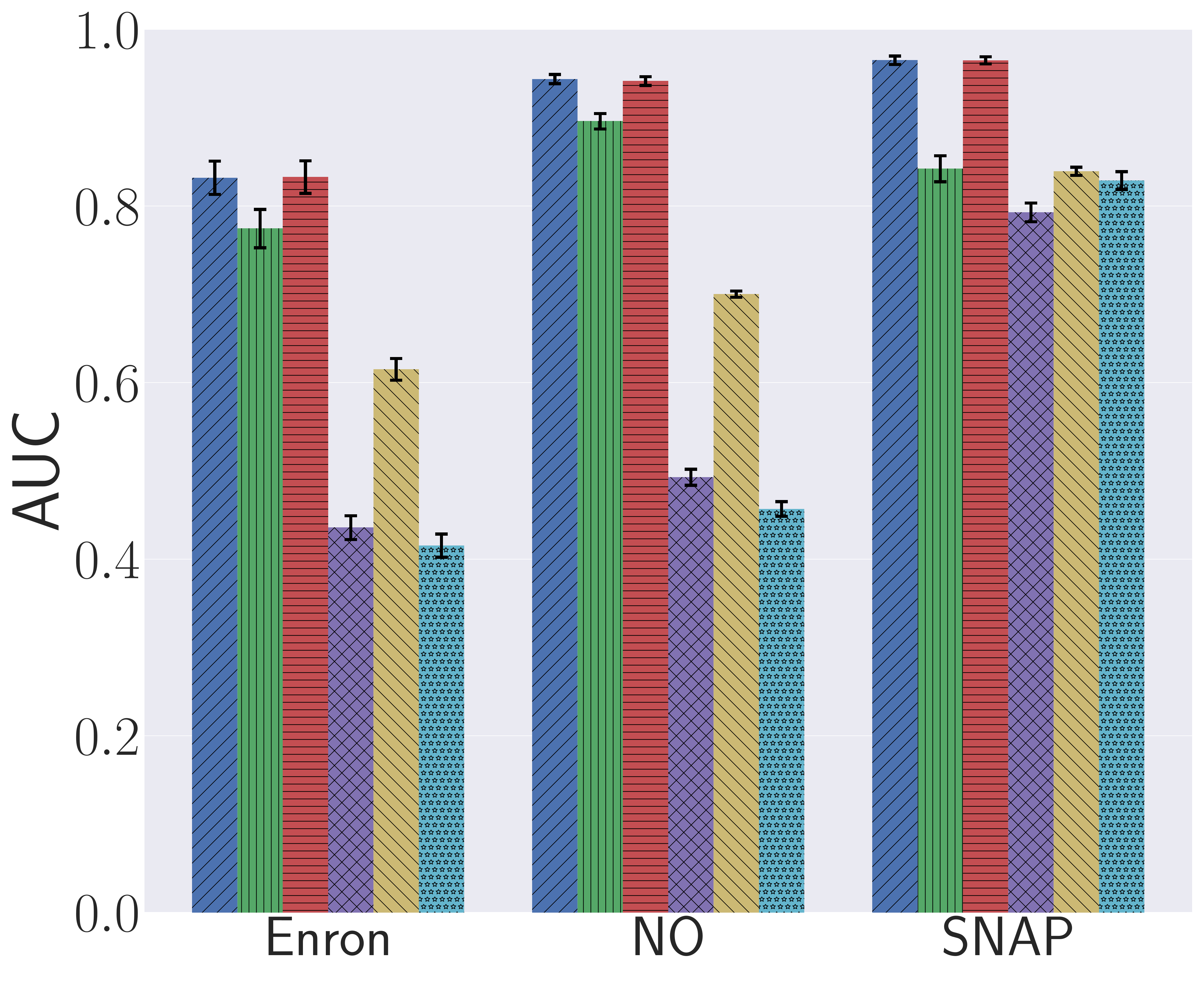}
\caption{SalaDP ($\epsilon=50$)}
\label{fig:sdp_75_auc}
\end{subfigure}
\begin{subfigure}{0.67\columnwidth}
\includegraphics[width=\columnwidth]{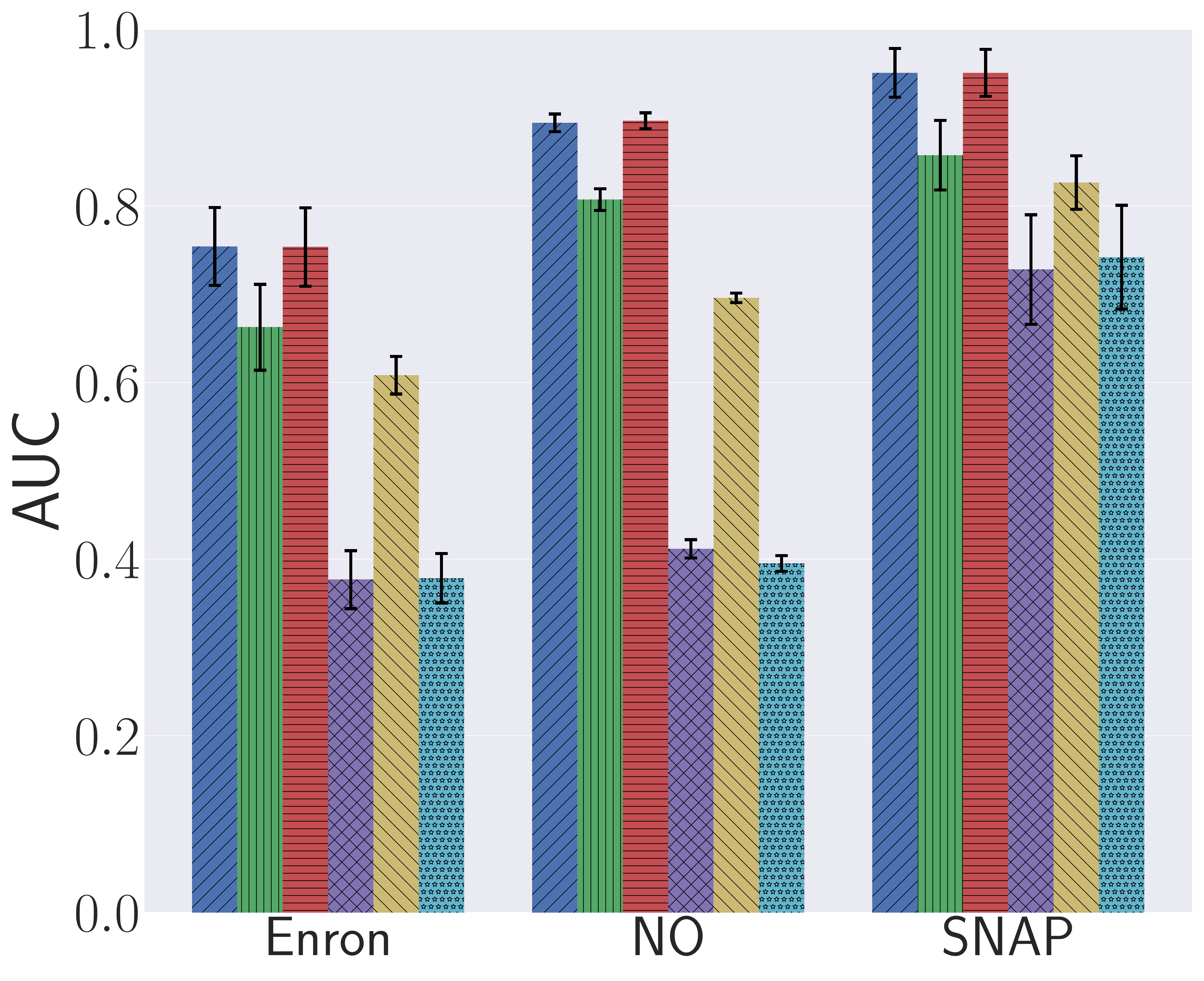}
\caption{SalaDP ($\epsilon=10$)}
\label{fig:sdp_50_auc}
\end{subfigure}
\caption{
[Higher is better] AUC scores for detecting fake edges for different datasets, 
structural proximity, distance metrics,
and anonymity levels ($k$ resp. $\epsilon$). 
The embedding approach clearly outperforms all three traditional structural proximity metrics. Moreover, cosine similarity performs best, only matched by Bray-Curtis distance.
}
\label{fig:auc}
\end{figure*} 

\section{Graph Recovery}
\label{sec:evalua}

In this section, we first evaluate the effectiveness of our edge plausibility metric
on differentiating fake edges from original ones 
without determining a decision threshold on the edge plausibility a priori.
Then, we present a method to automatically decide whether an edge is fake, 
which allows us to eventually recover the original graph.

\subsection{Experimental Setup}

\begin{table}[!t]
\centering
\caption{Statistics of the datasets.}
\label{table:dataset}
\begin{tabular}{l c c c}
\toprule
& Enron & NO & SNAP\\
\midrule
Number of users & 36,692 & 63,731 & 4,039\\
Number of edges & 183,831 & 817,090 & 88,234\\
Average degree & 10.020 & 25.642 & 43.691\\
Average clustering coefficient & 0.497 & 0.221 & 0.606\\
Number of triangles	& 727,044 &  3,501,542 & 1,612,010\\
\bottomrule
\end{tabular}
\end{table}

\mypara{Datasets}
We utilize three datasets for our experiments.
The first one, referred to as Enron, is a network of Email communications
in the Enron corporation.\footnote{\url{https://snap.stanford.edu/data/email-Enron.html}}
The second dataset (NO) is collected from Facebook users 
in the New Orleans area by Viswanath et al.~\cite{VMCG09}.
The third dataset (SNAP) by McAuley and Leskovec 
is obtained through a survey study~\cite{ML12}.
Note that Enron and NO are the two datasets 
used in the evaluation of SecGraph as well~\cite{JLMHB15}.
\autoref{table:dataset} presents some basic statistics of the three datasets.

\mypara{Baseline Models and Evaluation Metrics}
To demonstrate the effectiveness of our plausibility metric,
which is essentially a structural proximity metric,
we compare it with 
three classical structural proximity metrics,
namely, embeddedness (number of common friends), Jaccard index, and Adamic-Adar score~\cite{AA03}.
Their formal definition is as the following.
\[
\text{Embeddedness}: \vert\friend_{\ano}(\user) \cap \friend_{\ano}(\user') \vert
\]
\[
\text{Jaccard index}: \frac{\vert\friend_{\ano}(\user) \cap \friend_{\ano}(\user') \vert}{\vert\friend_{\ano}(\user) \cup \friend_{\ano}(\user') \vert}
\]
\[
\text{Adamic-Adar score}: \sum_{v\in \friend_{\ano}(\user) \cap \friend_{\ano}(\user')} \frac{1}{\log \vert \friend_{\ano}(v)\vert}
\]

Recall that cosine similarity is adopted
for measuring edge plausibility based on the users' vectors learned from graph embedding.
We also test two other vector similarity (distance) metrics,
namely the Euclidean distance and the Bray-Curtis distance,
defined as follows:
\[
\text{Euclidean}: \vert\vert f(\user) - f(\user') \vert\vert_2
\]
\[
\text{Bray-Curtis}: \frac{\sum_{i=1}^d \vert f(\user)_i-f(\user')_i \vert }{\sum_{i=1}^d \vert f(\user)_i+f(\user')_i \vert}
\]
Here, $f(\user)_i$ is the $i$-th element of vector $f(\user)$.

For evaluation metrics, 
we first use the AUC, which measures the area under the ROC curve.
The ROC curve projects the relation between
false-positive rate (on the x-axis) and true-positive rate (on the y-axis) 
over a series of thresholds for a given prediction task. 
A ROC curve closer to the top-left border of the plot 
(high true-positive rate for low false-positive rate), thus a larger AUC value, 
indicates higher prediction performance.
Morever, there exists a conventional standard 
to interpret AUC values:\footnote{\url{http://gim.unmc.edu/dxtests/roc3.htm}}
AUC = 0.5 is equivalent to random guessing, whereas an AUC
greater than 0.9 implies an excellent prediction.
Many recent works on assessing privacy risks have adopted AUC 
as the evaluation metric~\cite{BHPZ17,PTC18,HZHBTWB19,FLJLPR14,SZHBFB19,JSBZG19}.
We also make use of the F1 score for the method that automatically detects fake edges. 
Due to the randomness of the anonymization alogrithms, 
we repeat our experiments five times and report the average results.

\mypara{Parameters in Anonymization Mechanisms}
We rely on SecGraph to perform $k$-DA and SalaDP~\cite{JLMHB15}.
Each anonymization mechanism has its own privacy parameter.
For $k$-DA, we need to choose the value $k$, 
i.e., the minimal number of users sharing a certain degree 
for all possible degrees in $\Graph_{\ano}$.
Greater $k$ implies stronger privacy.
In our experiments, we choose $k$ to be 50, 75, and 100, respectively,
to explore different levels of privacy protection~\cite{JLMHB15}.
For SalaDP, the privacy parameter is $\epsilon$ 
which controls the noise added to the $dK$-2 series of $\Graph$: 
The smaller $\epsilon$  is, the higher its privacy provision is.
Following previous works~\cite{SZWZZ11,JLMHB15}, 
we experiment with three different $\epsilon$ values: 10, 50, and 100.
As stated before, both $k$-DA and SalaDP's principal operation
is adding fake edges to the original graph.
By running the two anonymization mechanisms on our three datasets,
we discover that this is indeed the case.
For instance, SalaDP ($\epsilon=10$)
adds 120\% more edges to the NO dataset,
while only deleting 1.7\% of the original edges.

\mypara{Hyperparameter Setting}
There are mainly three hyperparameters in the graph embedding phase:
walk length ($l$), walk times ($t$) and vector dimension ($d$).
For both $k$-DA and SalaDP, we choose $l=100$ and $t=80$.
Meanwhile, we set $d=128$ for $k$-DA and $d=512$ for SalaDP.
These values are selected through cross 
validation (see \autoref{subsec:hyper}).
For reproducibility purposes,
our source code will be made publicly available.

\begin{figure*}
\centering
\begin{subfigure}{0.67\columnwidth}
\includegraphics[width=\columnwidth]{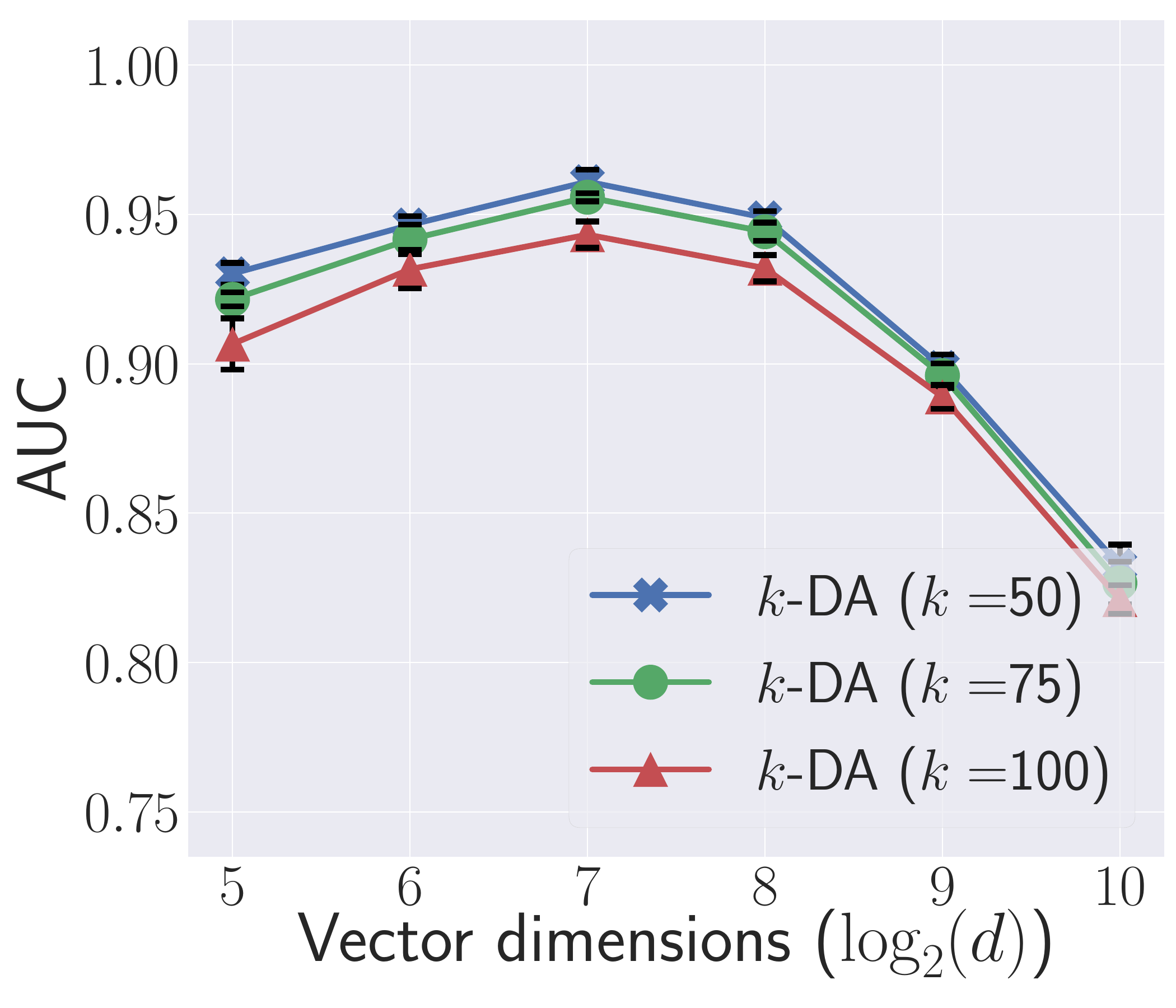}
\caption{Enron}
\label{fig:enron_kDa_num_features}
\end{subfigure}
\begin{subfigure}{0.67\columnwidth}
\includegraphics[width=\columnwidth]{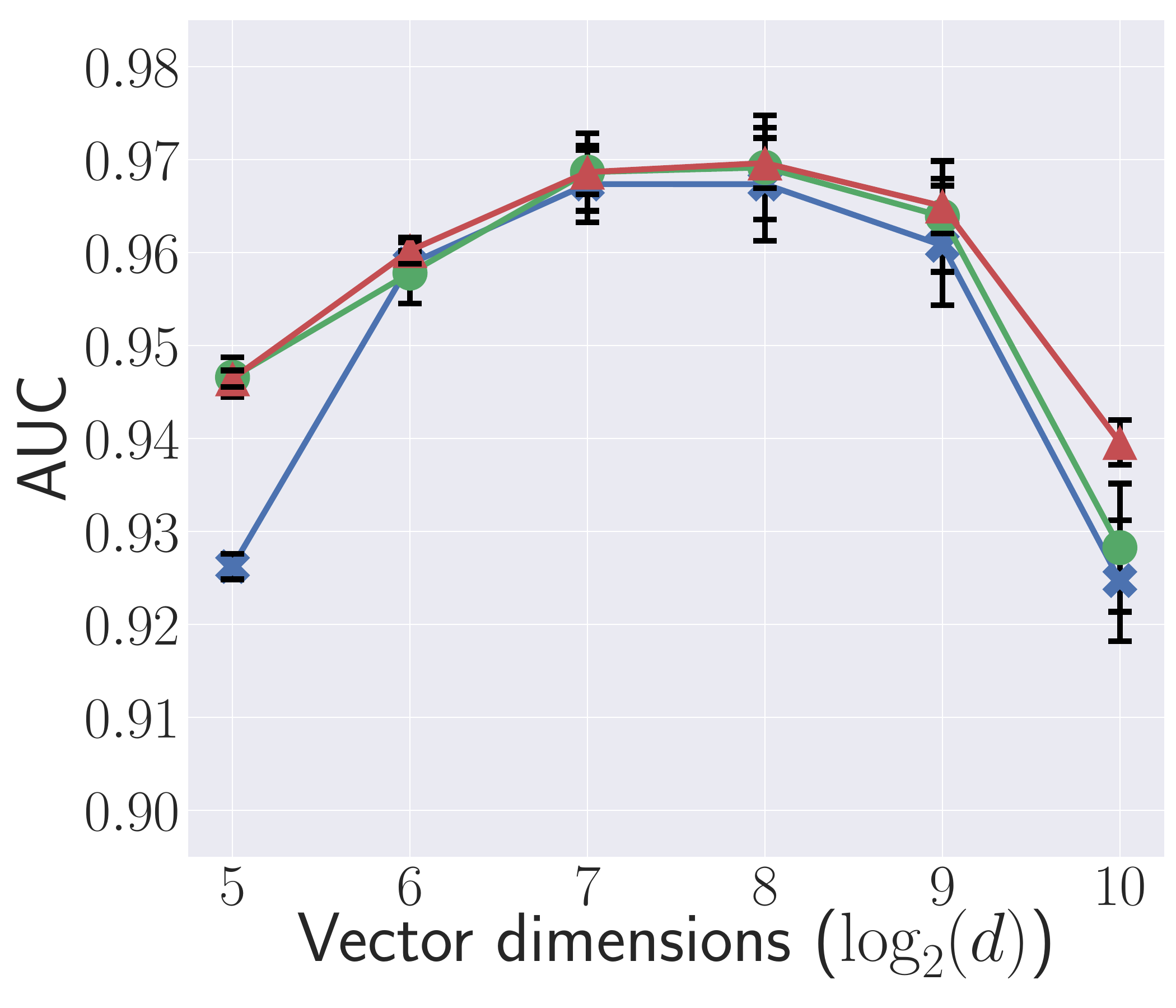}
\caption{NO}
\label{fig:facebook_kDa_num_features}
\end{subfigure}
\begin{subfigure}{0.67\columnwidth}
\includegraphics[width=\columnwidth]{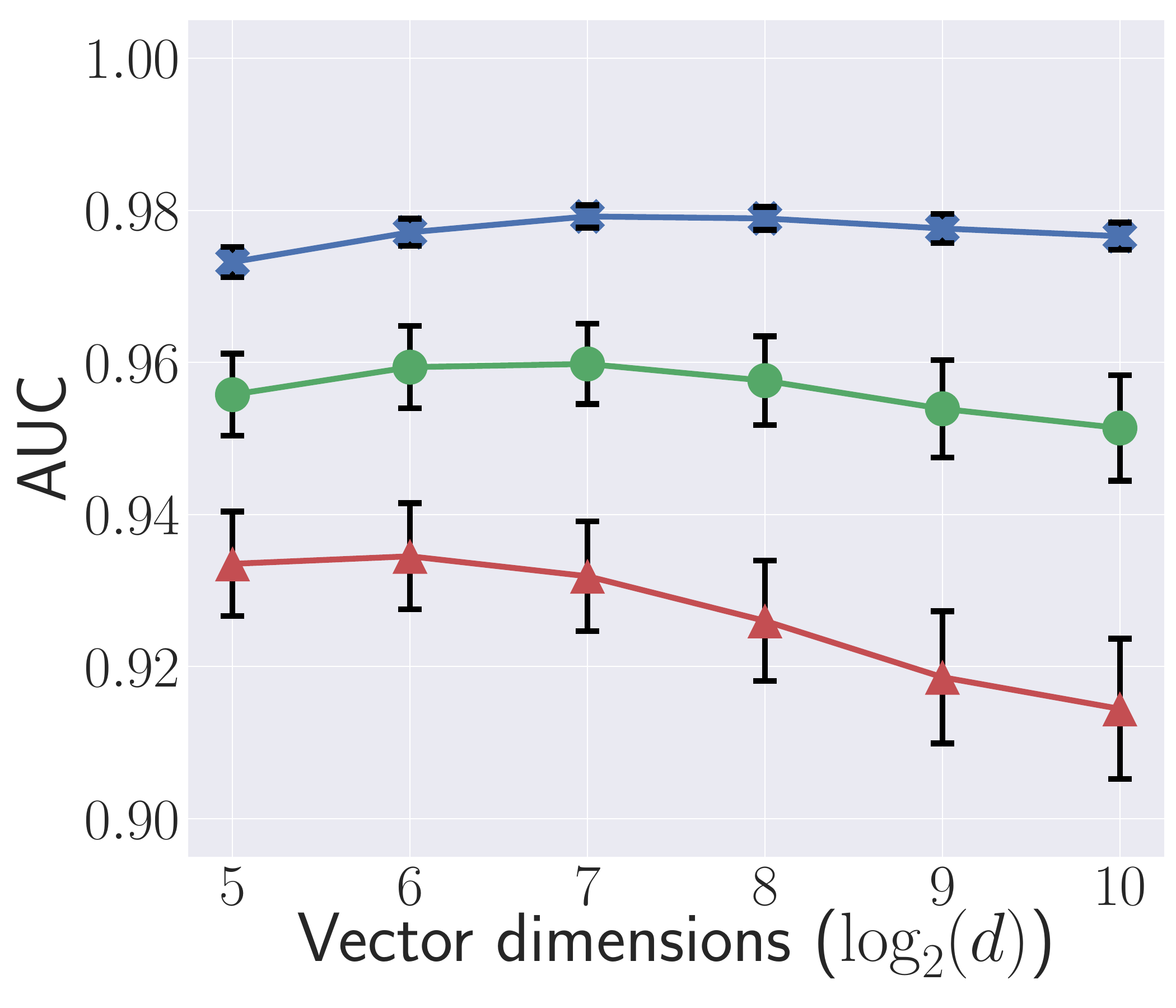}
\caption{SNAP}
\label{fig:snapfacebook_kDa_num_features}
\end{subfigure}
\caption{[Higher is better] Sensitivity of the AUC with respect to 
the embedding vector dimension  for $k$-DA-anonymized datasets.}
\label{fig:kDa_num_features}
\end{figure*} 

\begin{figure*}
\centering
\begin{subfigure}{0.67\columnwidth}
\includegraphics[width=\columnwidth]{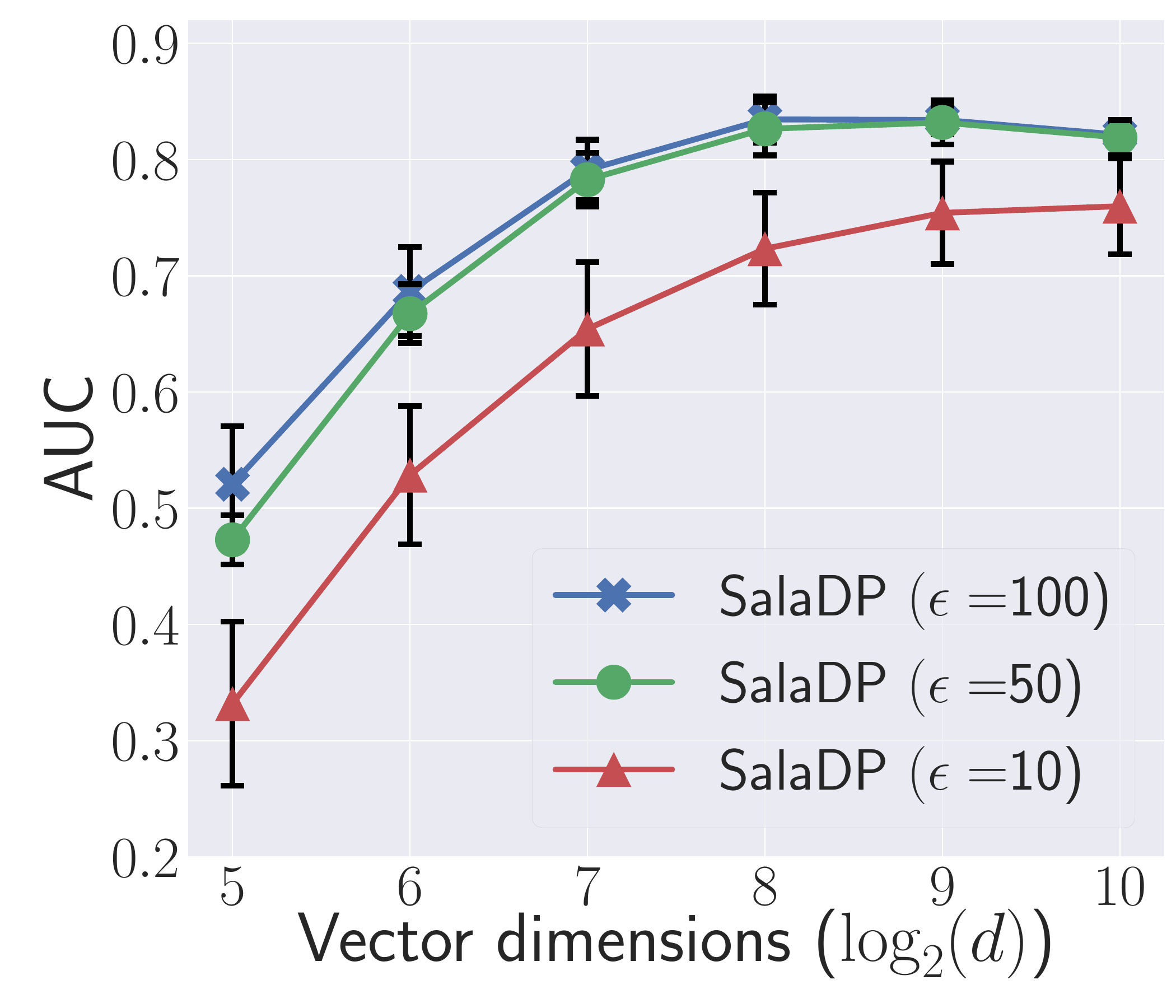}
\caption{Enron}
\label{fig:enron_sdp_num_features}
\end{subfigure}
\begin{subfigure}{0.67\columnwidth}
\includegraphics[width=\columnwidth]{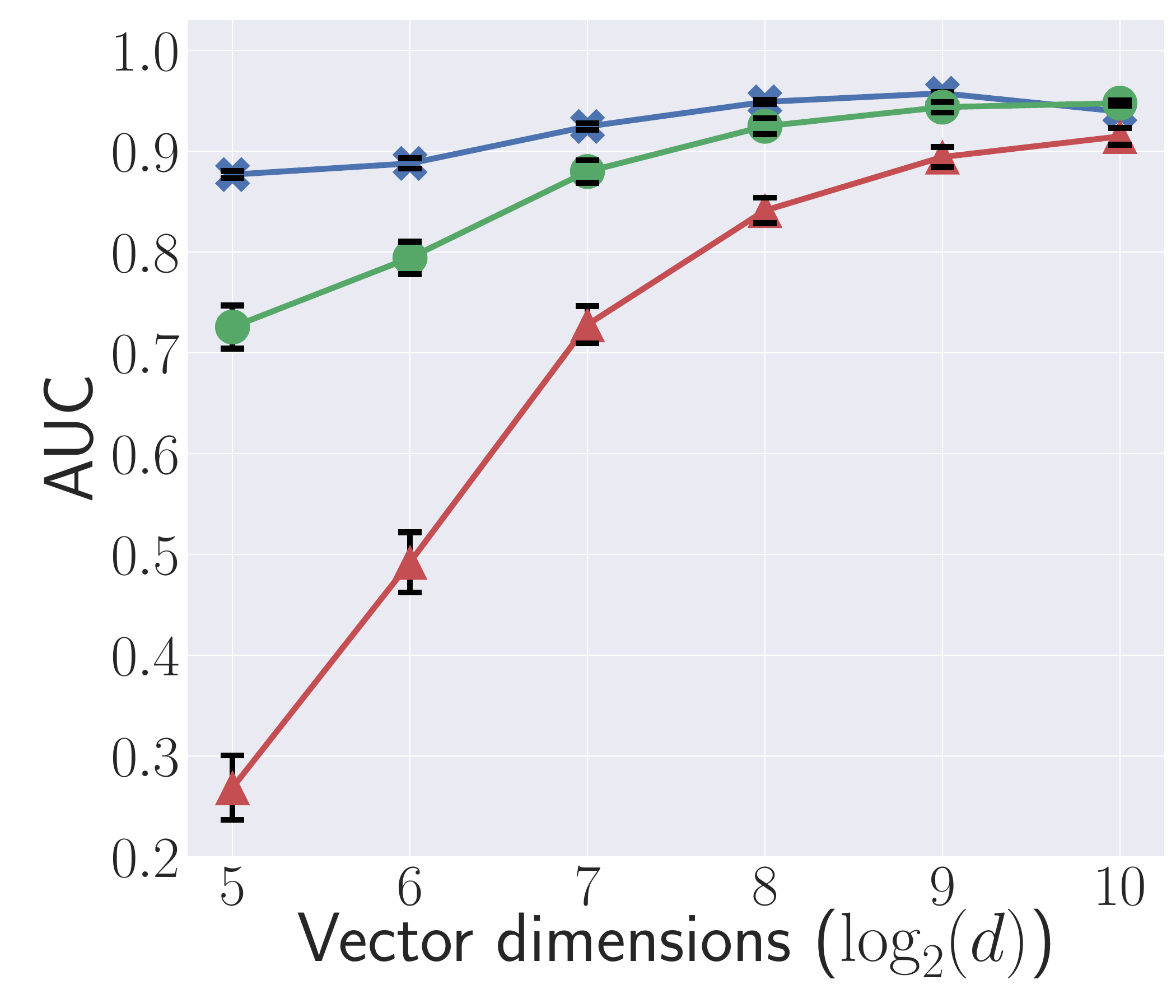}
\caption{NO}
\label{fig:facebook_sdp_num_features}
\end{subfigure}
\begin{subfigure}{0.67\columnwidth}
\includegraphics[width=\columnwidth]{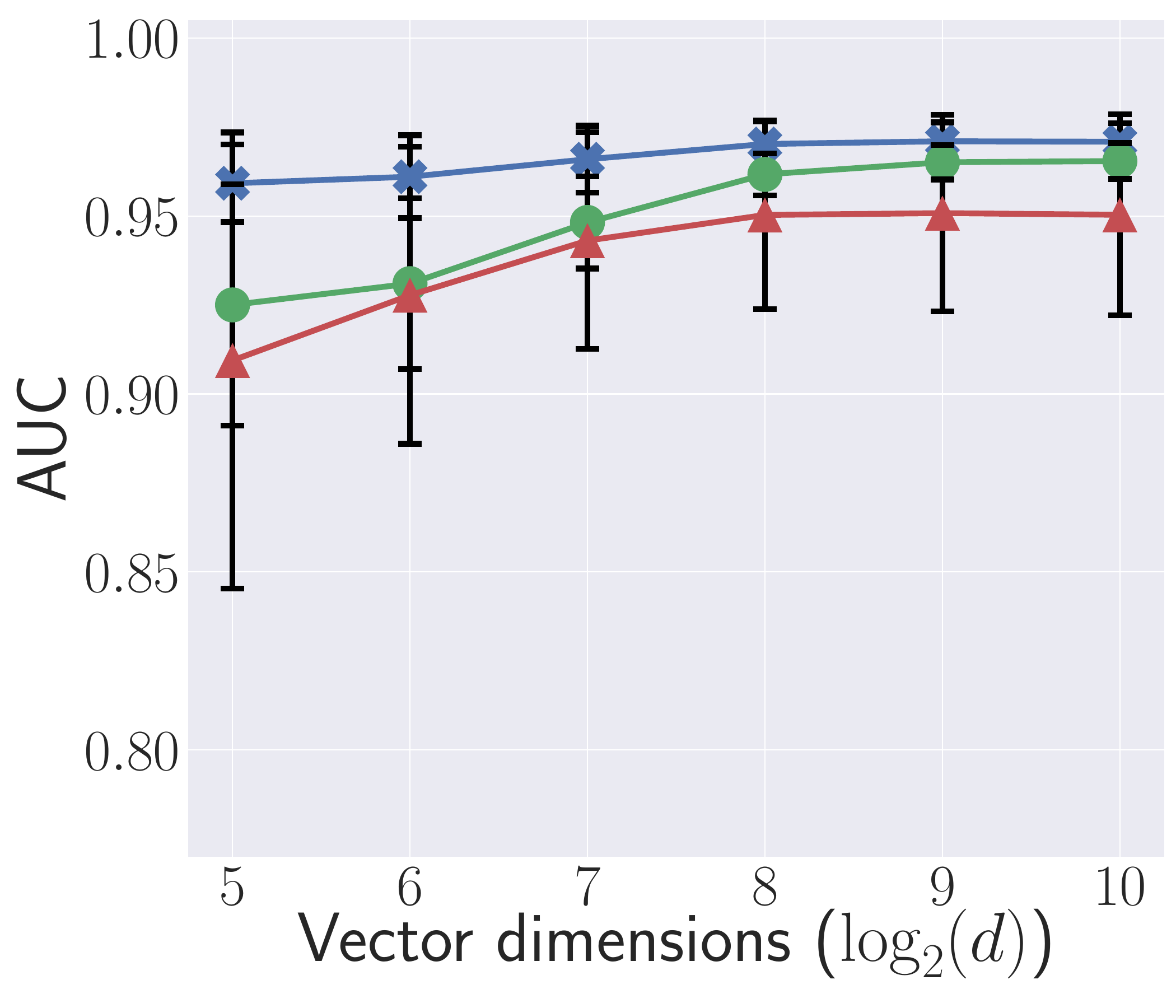}
\caption{SNAP}
\label{fig:snapfacebook_sdp_num_features}
\end{subfigure}
\caption{[Higher is better] Sensitivity of the AUC with respect to 
the embedding vector dimension for SalaDP-anonymized datasets.}
\label{fig:sdp_num_features}
\end{figure*} 

\subsection{Prediction Results}

\autoref{fig:auc} depicts the AUC values 
of using our edge plausibility metric (Cosine in \autoref{fig:auc})
to differentiate fake edges from original ones.
In most of the cases, we achieve excellent performance with AUC values above 0.95.
In particular, for the SalaDP-anonymized SNAP dataset ($\epsilon=100$),
the average AUC value is 0.971 (see \autoref{fig:sdp_100_auc}).
The only case where our edge plausibility 
does not achieve an excellent performance
is when applying SalaDP on the Enron dataset 
where the AUC values are between 0.76 and 0.83. 
However, we emphasize that for most of the classification tasks,
such AUC is already considered good.

We also notice that our method performs better against SalaDP on the SNAP dataset than the the other two. 
One reason is that SNAP has the highest number of average degrees (\autoref{table:dataset}),
which implies more diverse $dK$-2 series.
This results in SalaDP adding more fake edges on SNAP,
which leads to high performance of fake edge detection.
However, we do not observe a similar trend for $k$-DA-anonymized datasets.

The AUC values for other vector similarity (distance) metrics
are presented in \autoref{fig:auc} as well.
Cosine similarity performs slightly better than 
both Euclidean distance and Bray-Curtis distance
on $k$-DA-anonymized graphs.
On the other hand, for SalaDP-anonymized graphs, 
we can observe that cosine similarity
performs better than Euclidean distance (around 10\% performance gain),
while the performance of Bray-Curtis and cosine similarity is still very close.
This shows that cosine similarity (as well as Bray-Curtis distance) 
is a suitable choice for our edge plausibility metric.

\autoref{fig:auc} also shows that 
our edge plausibility significantly outperforms 
the traditional structural proximity metrics.
For instance, on the SalaDP-anonymized NO dataset ($\epsilon=50$),
our approach achieves 0.944 AUC 
while the result for the best performing structural proximity, 
i.e., Jaccard index, is  around 0.7.
It also appears that embeddedness outperforms the other two metrics
on $k$-DA-anonymized dataset in most of the cases,
while Jaccard index is rather effective for SalaDP. 

\subsection{Hyperparameter Sensitivity}
\label{subsec:hyper}

We study the influences of the three hyperparameters ($l$, $t$ and $d$) 
on the prediction performance.
Here, $l$ and $t$ are directly related to the size of the random walk traces,
which essentially decides the amount of data used for learning embedding vectors.
For both anonymization mechanisms,
we observe that increasing $l$ and $t$ improves the AUC values.
However, the increase is smaller when both of these values are above 60.
Therefore, we set $l=100$ and $t=80$.

Meanwhile, we observe interesting results for the vector dimension $d$:
different anonymization mechanisms have different optimal choices for $d$
(\autoref{fig:kDa_num_features} and \autoref{fig:sdp_num_features}).
It appears that when detecting fake edges on $k$-DA-anonymized graphs,
$d=128$ is a suitable choice for all datasets. 
On the other hand, for SalaDP,  
$d=512$ is able to achieve a stronger prediction.
We confirm that the vector dimension is indeed a subtle parameter,
as was observed in other data domains, 
such as biomedical data~\cite{BBHHKM16} and mobility data~\cite{BHPZ17}.
In conclusion, our default hyperparameter settings are suitable for our prediction task.

\begin{figure*}[!t]
\centering
\begin{subfigure}{0.67\columnwidth}
\includegraphics[width=\columnwidth]{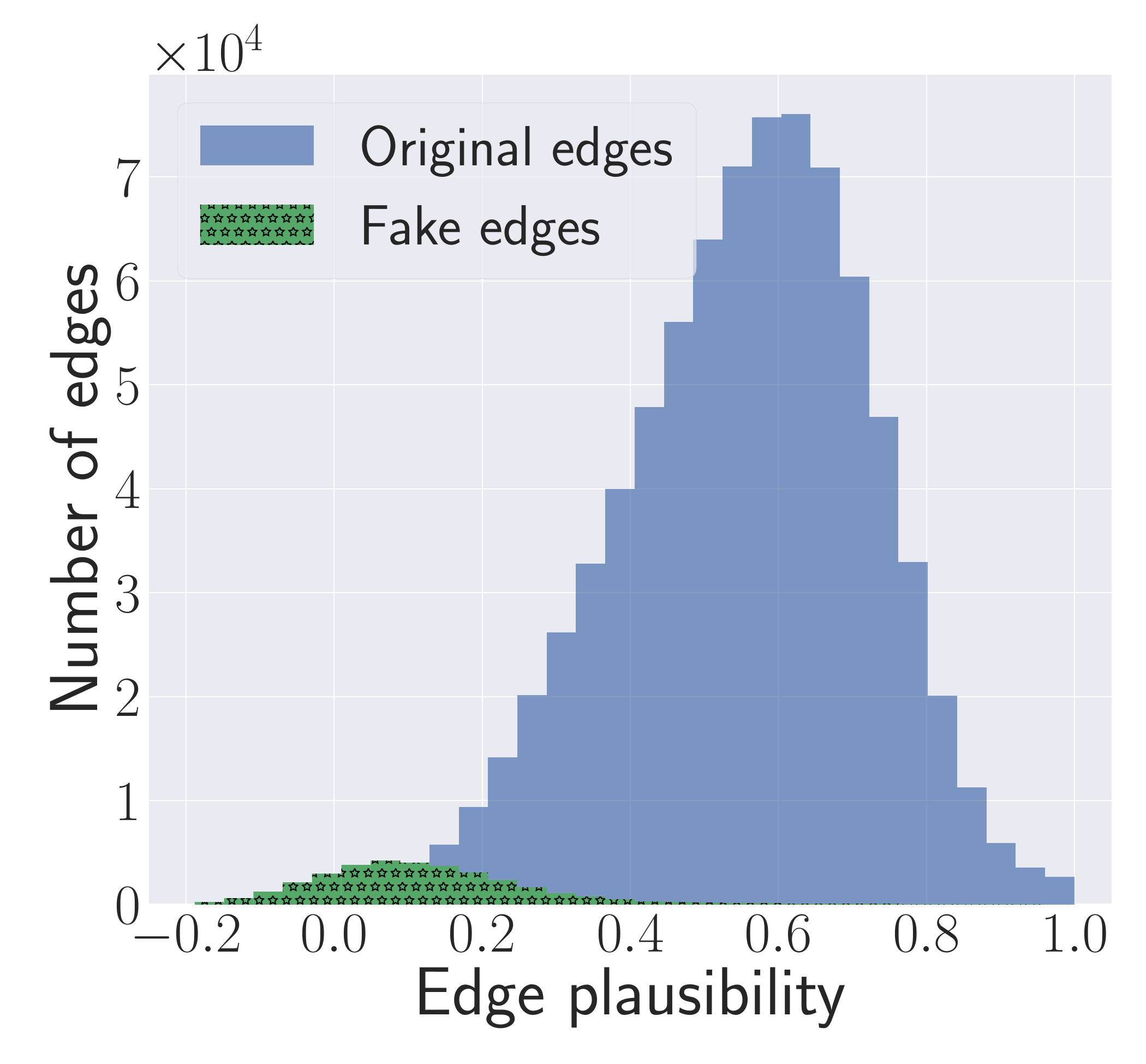}
\caption{$k$-DA ($k=50$)}
\label{fig:facebook_kDa_50_dist_hist}
\end{subfigure}
\begin{subfigure}{0.67\columnwidth}
\includegraphics[width=\columnwidth]{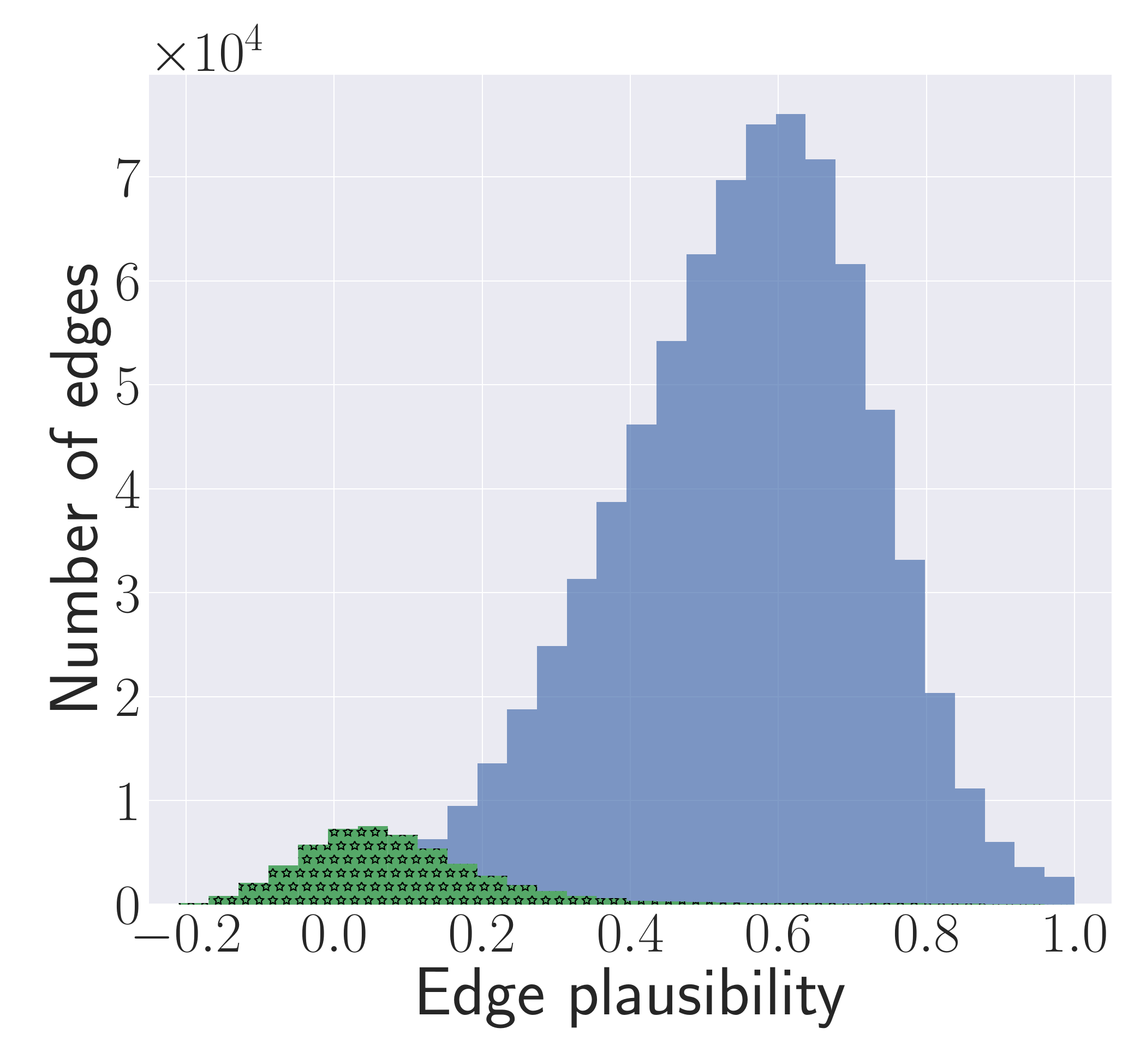}
\caption{$k$-DA ($k=75$)}
\label{fig:facebook_kDa_75_dist_hist}
\end{subfigure}
\\
\begin{subfigure}{0.67\columnwidth}
\includegraphics[width=\columnwidth]{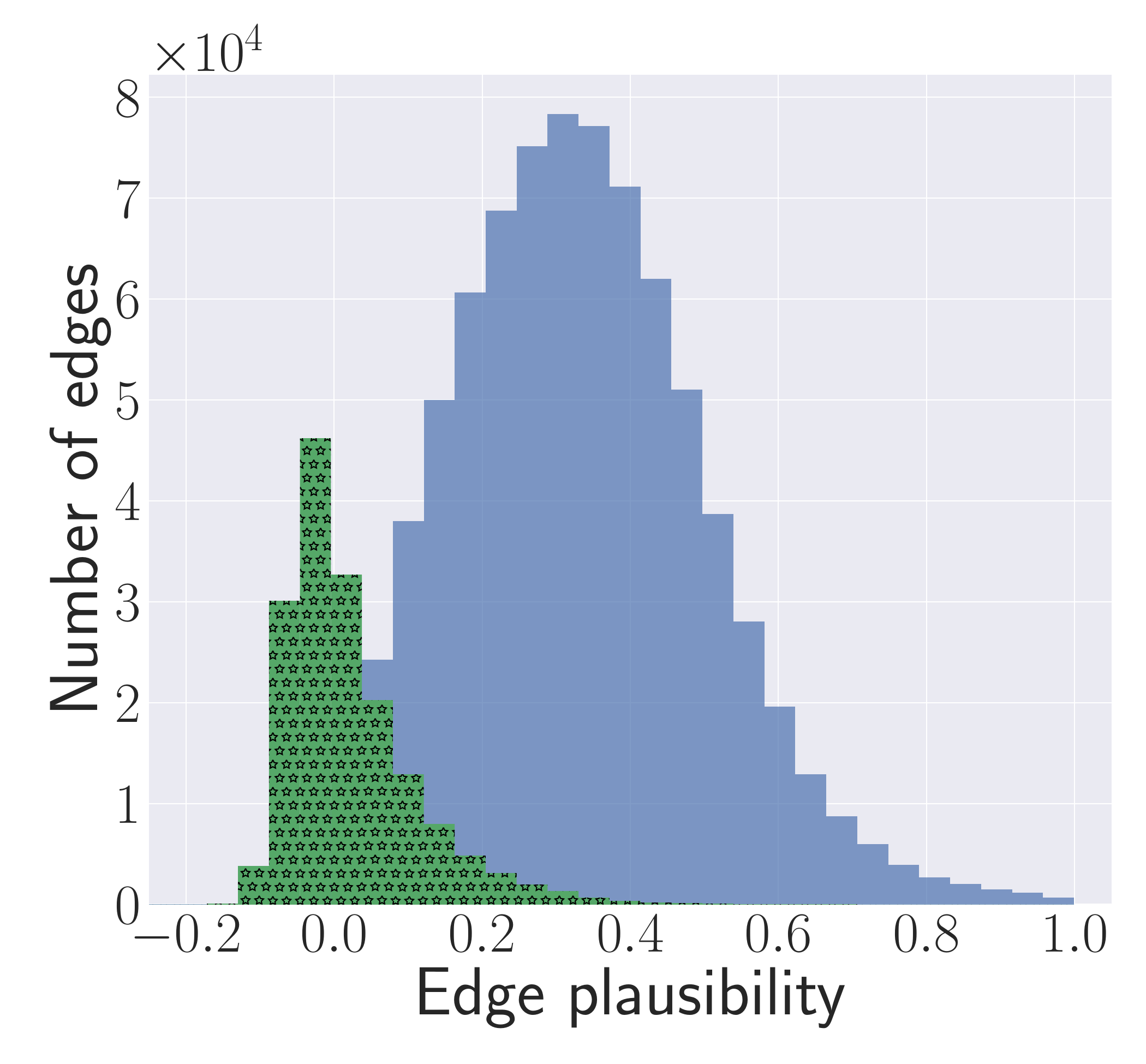}
\caption{SalaDP ($\epsilon=100$)}
\label{fig:facebook_sdp_100_dist_hist}
\end{subfigure}
\begin{subfigure}{0.67\columnwidth}
\includegraphics[width=\columnwidth]{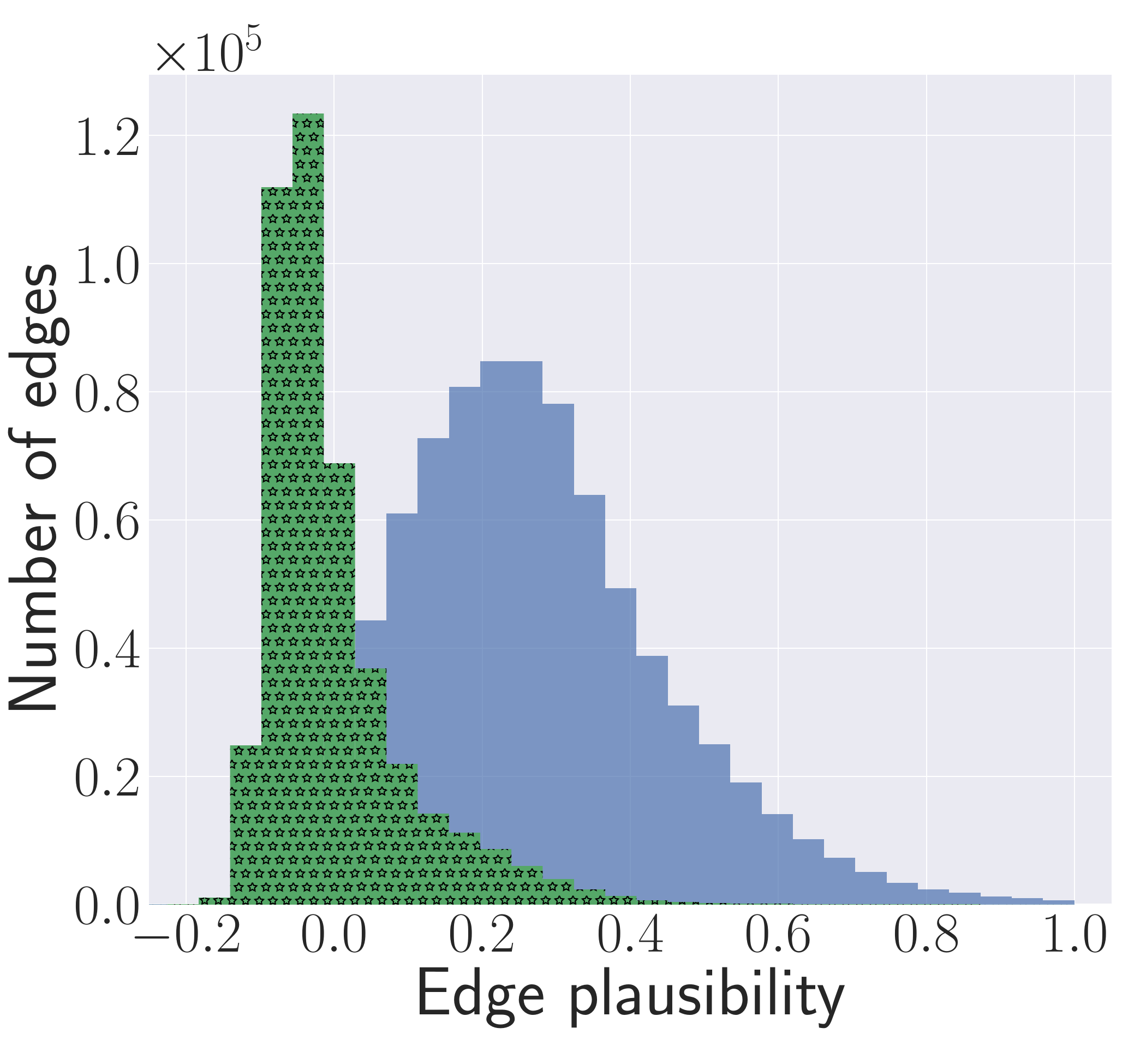}
\caption{SalaDP ($\epsilon=50$)}
\label{fig:facebook_sdp_50_dist_hist}
\end{subfigure}
\begin{subfigure}{0.67\columnwidth}
\includegraphics[width=\columnwidth]{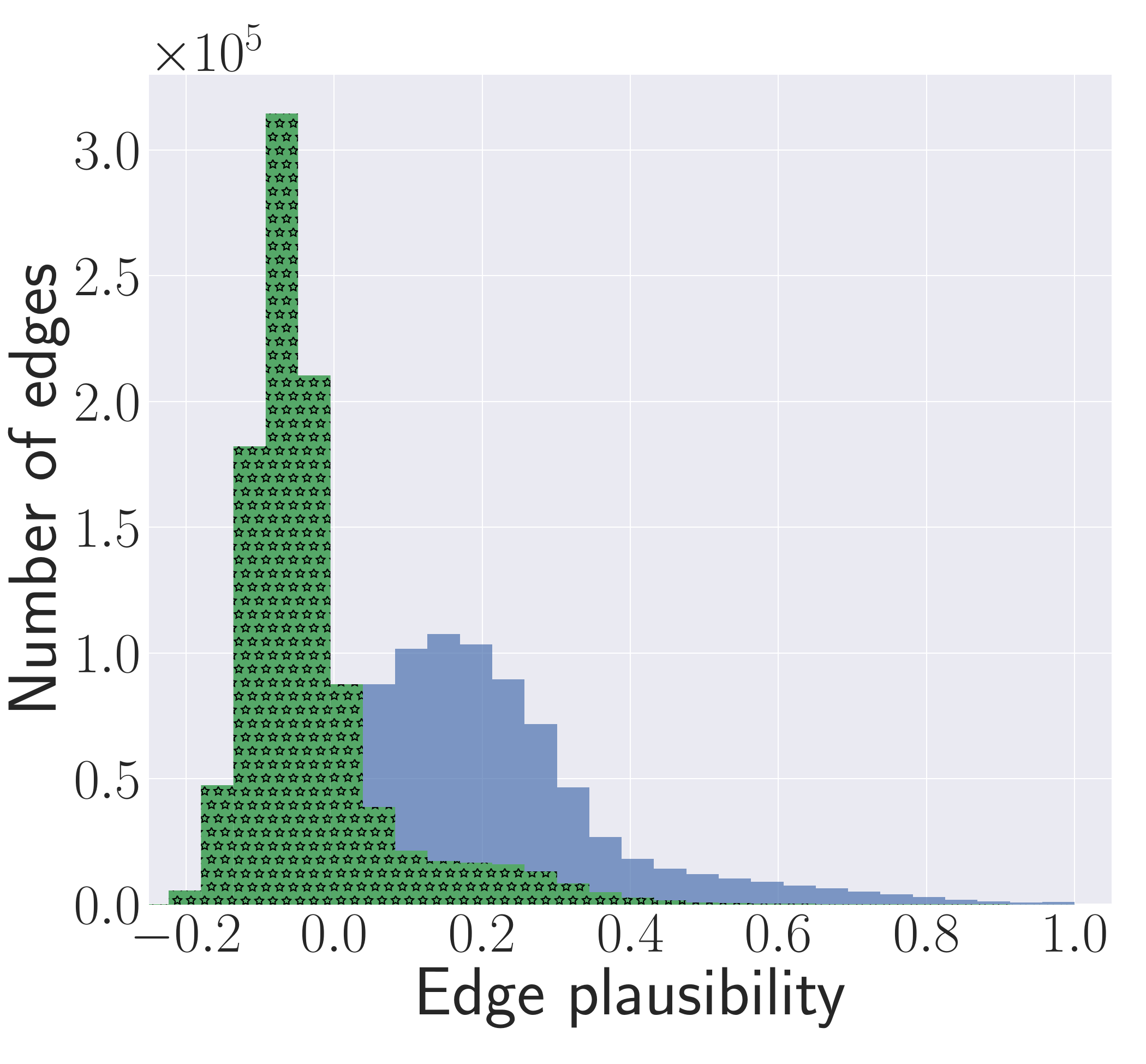}
\caption{SalaDP ($\epsilon=10$)}
\label{fig:facebook_sdp_10_dist_hist}
\end{subfigure}
\caption{Plausibility distributions of fake and original edges 
in the NO dataset anonymized by the two anonymization mechanisms.
The result for $k$-DA ($k=100$) is depicted in \autoref{fig:facebook_kDa_100_dist_hist}.
}
\label{fig:facebook_rest_dist_hist}
\end{figure*} 

\subsection{Optimizing Fake Edge Detection}

Next, we investigate how to concretely determine 
whether an edge in an anonymized graph is fake given its plausibility,
such that the adversary can recover the original graph from the anonymized one.

\autoref{fig:facebook_kDa_100_dist_hist} 
and \autoref{fig:facebook_rest_dist_hist}
depict the histograms of both fake and original edges' plausibility
in anonymized NO dataset (by both $k$-DA and SalaDP).
We see that both follow a Gaussian distribution
with different means and standard deviations.
Similar results are observed 
on Enron and SNAP datasets.

Given that the general population (plausibility of all edges)
consists of a mixture of two subpopulations (plausibility of fake and original edges)
with each one following a Gaussian distribution,
we can fit the general population with a Gaussian mixture model (GMM).
With the fitted GMM, we can obtain each edge's posterior
of being fake or original given its plausibility.
If the former is higher than the latter, 
then we predict the edge to be fake, 
effectively performing a maximum a posteriori (MAP) estimate.
This means GMM and MAP estimate provide us with 
a concrete threshold to determine whether an edge is fake given the observed data.

\mypara{Gaussian Mixture Model}
To formally define our GMM,
we first introduce two random variables: $\Fake$ and $\Proxi$.
$\Fake$ represents whether an edge is 
original ($\Fake=0$) or fake ($\Fake=1$),
while $\Proxi$ represents the plausibility of an edge.
The probability density function of our GMM is formally defined as:
\[
p(\Proxi =\proxi_{\ano}(\user, \user'))=\sum_{i \in \{0, 1\}} \weight_i \mathit{N}(\proxi_{\ano}(\user, \user')\vert \mu_i, \sigma_i).
\]
The GMM is parametrized by 6 parameters:
$\weight_0$, $\mu_0$, $\sigma_0$, $\weight_1$, $\mu_1$ and $\sigma_1$.
Here, $\weight_0$ ($\weight_1$) is the prior probability of an edge being original (fake),
i.e., $\weight_0 = P(\Fake=0)$ ($\weight_1 = P(\Fake=1)$).
The other 4 parameters are related to the two Gaussian distributions
for edge plausibility:
$\mathit{N}(\proxi_{\ano}(\user, \user')\vert \mu_i, \sigma_i)$ for $i \in \{0, 1\}$
is the density function of the Gaussian distribution:
\[
\frac{1}{\sqrt{2\pi \sigma_i^2}}\exp(-\frac{1}{2\sigma_i^2}(\proxi_{\ano}(\user, \user')-\mu_i)).
\]

\mypara{Parameter Learning}
To learn the 6 parameters of the GMM,
we adopt the expectation maximization (EM) algorithm,
which consists of two steps, i.e., the expectation (E) step and the maximization (M) step.
The E-step calculates, for each edge in $\Graph_{\ano}$,
its posterior probability of being fake or original given its plausibility value.
The M-step updates all the 6 parameters based on the probabilities calculated
from the E-step following maximum likelihood estimation.
The learning process iterates over the two steps until convergence.
Here, convergence means that the log-likelihoods of two consecutive iterations 
differ less than a given threshold (we set it to 0.001 in our experiments).
In addition, the initial values of the 6 parameters are set randomly.

\mypara{Fake Edge Detection}
After the GMM has been learned,
we compute for each edge $\{\user, \user'\}$ its posterior probabilities
of being original and fake:
\[
P(\Fake=0 \vert \proxi_{\ano}(\user, \user'))=
\frac{ \weight_0 \mathit{N}(\proxi_{\ano}(\user, \user')\vert \mu_0, \sigma_0)}
{\sum\limits_{i\in\{0, 1\}} \weight_i \mathit{N}(\proxi_{\ano}(\user, \user')\vert \mu_i, \sigma_i)}
\]
\[
P(\Fake=1 \vert \proxi_{\ano}(\user, \user'))=
\frac{ \weight_1 \mathit{N}(\proxi_{\ano}(\user, \user')\vert \mu_1, \sigma_1)}
{\sum\limits_{i\in\{0, 1\}} \weight_i \mathit{N}(\proxi_{\ano}(\user, \user')\vert \mu_i, \sigma_i)}
\]
and pick the one that is maximum (MAP estimate):
If $P(\Fake=1 \vert \proxi_{\ano}(\user, \user'))>P(\Fake=0 \vert \proxi_{\ano}(\user, \user'))$,
we predict $\{\user, \user'\}$ to be fake, and vice versa.

In the end, we delete all the predicted fake edges,
and obtain the recovered graph $\Graph_{\rec}$.

\begin{table}[!t]
\centering
\caption{
[Higher is better] F1 scores for detecting fake edges 
using GMM and MAP estimate for both $k$-DA and SalaDP
on three different datasets.}
\label{table:f1}
\begin{tabular}{l  c c c }
\toprule
& Enron & NO & SNAP\\
\midrule
$k$-DA ($k=50$) & 0.792 & 0.642 & 0.857\\
$k$-DA ($k=75$) & 0.796 & 0.710 & 0.869\\
$k$-DA ($k=100$) & 0.812 & 0.761 & 0.881\\
\midrule
SalaDP ($\epsilon=100$) & 0.672 & 0.712 & 0.853 \\
SalaDP ($\epsilon=50$) & 0.750 & 0.723 & 0.835 \\
SalaDP ($\epsilon=10$) & 0.819 & 0.876 & 0.802\\
\bottomrule
\end{tabular}
\end{table}

\mypara{Results}
We train GMMs under both anonymization mechanisms for all the datasets.
\autoref{table:f1} presents the results.
We first observe that, in most of the cases,
our approach achieves a strong prediction,
e.g., for the SalaDP-anonymized NO dataset ($\epsilon=10$),
the F1 score is 0.876.
For our worst prediction on SalaDP-anonymized Enron dataset ($\epsilon=100$),
the F1 score is still approaching 0.7.
Another interesting observation is that
when the privacy level increases, i.e., higher $k$ or lower $\epsilon$,
our prediction performance increases in most of the cases.
This can be explained by the fact that higher privacy levels
lead to more fake edges being added.
\section{Privacy Loss}
\label{sec:prideg}

As fake edges help an anonymized graph $\Graph_{\ano}$ satisfy certain privacy guarantees,
we expect that, by obtaining the recovered graph $\Graph_{\rec}$ from $\Graph_{\ano}$,
these guarantees will be violated.
In this section, we first define two metrics 
tailored to each anonymization mechanism 
for quantifying the privacy loss due to our graph recovery attack.
Then, we present the corresponding evaluation results.

\subsection{Privacy Loss Measurement}

\mypara{$k$-DA}
$k$-DA assumes that the adversary 
only has knowledge of her targets' degrees
and uses this knowledge to re-identify them.
This means that, if the users' degrees in $\Graph_{\rec}$
are more similar to those in $\Graph$ compared to $\Graph_{\ano}$,
then the adversary is more likely to achieve her goal.
Therefore, we propose to compute users' 
average degree difference between $\Graph_{\ano}$ and $\Graph$,
as well as between $\Graph_{\rec}$ and $\Graph$,
to measure the privacy loss 
caused by our graph recovery.
Formally, we define users' average degree difference between
$\Graph_{\ano}$ and $\Graph$ as:
\[
\degdiff_{\ano}=
\frac{\sum\limits_{\user \in \User}\vert \vert\friend(\user)\vert -\vert\friend_{\ano}(\user)\vert \vert}{\vert \User \vert}
\]
and define users' average degree difference between $\Graph$ and $\Graph_{\rec}$
($\degdiff_{\rec}$) accordingly.

Note that our approach also 
deletes some original edges when recovering $\Graph_{\rec}$ (false positives).
Therefore, if the adversary relies on the users' exact degrees (as assumed in $k$-DA) 
to de-anonymize them, she might fail.
However, a sophisticated adversary can apply extra heuristics,
such as tolerating some degree differences for finding her targets.
In this case, $\degdiff_{\rec}$ being smaller than $\degdiff_{\ano}$
can still provide the adversary with a better chance to achieve her goal.

\mypara{SalaDP}
To quantify the privacy loss for SalaDP, 
we consider the noise added to the $dK$-2 series of the original graph $\Graph$.
Formally, the $dK$-2 series of $\Graph$, denoted by $\dk(\Graph)$, is a set
with each element $\dkelement_{i, j}(\Graph)$ in $\dk(\Graph)$
representing the number of edges that connect users of degrees $i$ and $j$ in $\Graph$.
Formally, $\dkelement_{i, j}(\Graph)$ is defined as:
\[
\dkelement_{i, j}(\Graph) =  \vert\{ \{\user, \user'\}\vert \{\user, \user'\} \in \Edge \wedge \vert \friend(\user)\vert=i 
\wedge \vert \friend(\user')\vert =j \} \vert.
\]
Accordingly, $\dkelement_{i, j}(\Graph_{\ano})$ and $\dkelement_{i, j}(\Graph_{\rec})$
represent the corresponding numbers in $\Graph_{\ano}$ and $\Graph_{\rec}$.
Then, we use $\noise_{i, j}(\Graph, \Graph_{\ano})=\dkelement_{i, j}(\Graph_{\ano})-\dkelement_{i, j}(\Graph)$
to denote the noise added to $\dkelement_{i, j}(\Graph)$ 
when transforming $\Graph$ to $\Graph_{\ano}$,
and $\noise_{i, j}(\Graph, \Graph_{\rec})=\dkelement_{i, j}(\Graph_{\rec})-\dkelement_{i, j}(\Graph)$
to represent the (lower) noise caused by our graph recovery attack. 
Since SalaDP is a statistical mechanism, 
we sample 100 anonymized graphs $\{\Graph_\ano^{t}\}_{t=1}^{100}$ 
by applying SalaDP to $\Graph$ 100 times
and produce 100 noise samples $\{\noise_{i,j}(\Graph,\Graph_\ano^{t})\}_{t=1}^{100}$ 
for each element in $\dk(\Graph)$.

We define two metrics for quantifying the privacy loss 
due to graph recovery for SalaDP.
In the first metric, we compare the difference of average noise
added to $\dk(\Graph)$ before (by SalaDP) and after graph recovery.
Concretely, for each $\dkelement_{i,j}(\Graph)$ in $\dk(\Graph)$,
we first calculate the average absolute noise added to $\dkelement_{i,j}(\Graph)$,
denoted by $\noisesample_{i, j}(\Graph, \Graph_{\ano})$,
over the 100 SalaDP graph samples described above,
i.e., 
\[
\noisesample_{i, j}(\Graph, \Graph_{\ano})=\frac{\sum_{t=1}^{100}\vert \noise_{i,j}(\Graph,\Graph_\ano^t)\vert }{100}.
\]
Then, we compute the average noise over the whole graph as:
\[
\noise_\ano = \frac{1}{\vert\dk(\Graph)\vert} \sum_{\dkelement_{i, j}(\Graph) \in \dk(\Graph)} \noisesample_{i,j}(\Graph, \Graph_{\ano}).
\]
We analogously compute the average added noise $\noise_\rec$ after our graph recovery attack.

For the second approach, we consider the uncertainty introduced by the added noise.
McGregor et al. explore the connection 
between privacy and the uncertainty of the output 
produced by differential privacy mechanisms~\cite{MMPRTV10}. 
In general, higher uncertainty implies stronger privacy. 
We measure the uncertainty of noise added by SalaDP through estimating its empirical entropy. 
To this end, we calculate the Shannon entropy over the frequencies 
of elements in $\{\noise_{i,j}(\Graph,\Graph_\ano^{t})\}_{t=1}^{100}$ 
(the 100 noise samples described above),
denoted by $\entropy_{i, j}(\Graph, \Graph_{\ano})$.
Then, we compute the average entropy of the noise as follows:
\[
\entropy_\ano = \frac{1}{\vert\dk(\Graph) \vert } \sum_{\dkelement_{i, j}(\Graph) \in \dk(\Graph)} \entropy_{i,j}(\Graph,\Graph_{\ano}).
\]
We compute the average entropy 
after our graph recovery similarly and denote it as $\entropy_{\rec}$.

\subsection{Evaluation}

\mypara{$k$-DA} \autoref{table:degseq}
presents the results of the users' degree differences.
In all cases, $\degdiff_{\rec}$ is smaller than $\degdiff_{\ano}$.
This indicates that the adversary 
has a better chance to identify her targets from $\Graph_{\rec}$ than from $\Graph_{\ano}$, 
and demonstrates that our attack clearly decreases the privacy provided by $k$-DA.
It also appears that our graph recovery gains least benefits for the adversary on the NO dataset,
where $\degdiff_{\rec}$ is closer to $\degdiff_{\ano}$.
This is essentially due to the fact that the original NO dataset 
already preserves a high $k$-degree anonymity. 

\mypara{SalaDP}
\autoref{table:noise} presents the average noise 
added to the $dK$-2 series of the original graph
with respect to the anonymized and recovered graphs. 
We observe that, in all cases, $\noise_\rec$ is smaller than $\noise_\ano$ 
showing that our recovery attack reduces the average noise for SalaDP. 
We also observe that the relative reduction of the average noise 
with our graph recovery in general decreases when increasing $\epsilon$: 
The added noise is already much smaller for larger $\epsilon$
and cannot be further reduced.

\autoref{table:entropy} presents the average entropy of the noise 
added to the $dK$-2 series of the original graph after applying SalaDP 
and after the graph recovery attack. 
Note that, while one would expect higher entropy for smaller values of $\epsilon$, 
this does not hold true in practice 
because the SalaDP mechanism is not necessarily optimal 
with respect to the added uncertainty. 
Still, across all values of $\epsilon$ and all the datasets,
we can observe a reduction of the empirical entropy, 
and therefore a reduction of the privacy provision. 
The relative reduction, however, varies between the values of $\epsilon$ and, 
as for the average noise above, between the datasets. 

\begin{table}[!t]
\centering
\caption{
Differences in average degree between the original graph ($\Graph$),
the $k$-DA anonymized graph ($\Graph_{\ano}$) and our recovered graph ($\Graph_{\rec}$).
}
\label{table:degseq}
\setlength{\tabcolsep}{5pt}
\begin{tabular}{l  c  c  c  c  c  c }
\toprule
& \multicolumn{2}{c}{Enron} & \multicolumn{2}{c}{NO} & \multicolumn{2}{c}{SNAP}\\
\midrule
& $\degdiff_{\rec}$ & $\degdiff_{\ano}$ & $\degdiff_{\rec}$ & $\degdiff_{\ano}$ & $\degdiff_{\rec}$ & $\degdiff_{\ano}$\\
\midrule
$k$-DA ($k=50$) & 0.990&1.222&0.499&0.541&6.589&8.216\\
$k$-DA ($k=75$) & 1.367&1.705&0.752&0.875&8.815&11.755\\
$k$-DA ($k=100$) & 2.019&2.377&1.035&1.231&11.565&16.018\\
\bottomrule
\end{tabular}
\end{table}

\begin{table}[!t]
\centering
\caption{
Differences in average noise on the original graph $\Graph$'s $dK$-2 series
by SalaDP ($\noise_{\ano}$) and by our graph recovery attack ($\noise_{\rec}$).
}
\label{table:noise}
\setlength{\tabcolsep}{5pt}
\begin{tabular}{l c c c c c c}
\toprule
& \multicolumn{2}{c}{Enron} & \multicolumn{2}{c}{NO} & \multicolumn{2}{c}{SNAP}\\
\midrule
& $\noise_\rec$ & $\noise_\ano$ & $\noise_\rec$ & $\noise_\ano$ & $\noise_\rec$ & $\noise_\ano$\\
\midrule
SalaDP ($\epsilon=100$) & 4.432 & 5.282 & 6.048 & 6.415 & 3.422 & 4.018\\
SalaDP ($\epsilon=50$) & 4.224 & 7.121 & 7.731 &  9.471 & 3.489 & 4.445\\
SalaDP ($\epsilon=10$) & 4.958 & 12.004 & 7.982 & 16.033 & 3.672 & 5.690\\
\bottomrule
\end{tabular}
\end{table}

\begin{table}[!t]
\centering
\caption{
Differences in average entropy of the noise on the original graph $\Graph$'s $dK$-2 series
by SalaDP ($\entropy_{\ano}$) and by our graph recovery attack ($\entropy_{\rec}$).
}
\label{table:entropy}
\setlength{\tabcolsep}{5pt}
\begin{tabular}{l c c c c c c }
\toprule
& \multicolumn{2}{c}{Enron} & \multicolumn{2}{c}{NO} & \multicolumn{2}{c}{SNAP}\\
\midrule
& $\entropy_\rec$ & $\entropy_\ano$ & $\entropy_\rec$ & $\entropy_\ano$ & $\entropy_\rec$ & $\entropy_\ano$\\
\midrule
SalaDP ($\epsilon=100$) & 0.180 & 2.029 & 1.243 & 2.515 & 1.999 & 2.209\\
SalaDP ($\epsilon=50$) & 0.556 & 1.865 & 1.754 &  2.852 & 2.000 & 2.238\\
SalaDP ($\epsilon=10$) & 1.095 & 1.381 & 2.275 & 3.112 & 1.926 & 2.022\\
\bottomrule
\end{tabular}
\end{table}

For now, it seems unclear how these various factors
impact the relative reduction of empirical entropy. 
Analyzing the impact of these parameters 
on the relative reduction of empirical entropy 
could provide further insights into the recoverability of anonymized graphs. 
Such work is, however, orthogonal to the work 
presented in this paper and could be an interesting direction for our future work.  

Note that differential privacy guarantees are theoretically not violated since differential privacy 
is, by definition, closed under post-processing~\cite{DR14}.
However, despite these formal semantic guarantees are still valid, 
we demonstrate that our recovery attack can,
without additional knowledge about the data, 
reduce the magnitude of the actual noise put in place 
to perturb the original graph by exploiting the graph structure. 
This demonstrates that, by simply looking at the sanitized data, 
we can concretely jeopardize the anonymity of the graph.

\mypara{Graph De-anonymization}
We also compare the performance of the graph de-anonymization attack 
designed by Narayanan and Shmatikov~\cite{NS09},
referred to as the NS-attack,
on both anonymized and recovered social graphs. 
Our experiments show that, contrary to what one might initially expect, 
graph recovery does not improve 
the performance of the graph de-anonymization significantly.
Our explanation is that the NS-attack
assumes a much stronger adversary model,
such as an auxiliary graph with seed nodes already de-anonymized (see \autoref{sec:fix}).
Moreover, Ji et al.\ show that, in many cases, 
the NS-attack even performs better 
on the anonymized graph than on the original graph~\cite{JLMHB15}.
\section{Enhancing Graph Anonymization}
\label{sec:fix}

In this section, we take the first step towards
enhancing the existing graph anonymization mechanisms. 
We start by presenting our methodology, 
then evaluate the performance of fake edge detection as well as graph utility 
with the enhanced mechanisms.
In the end, we study our new anonymized graphs' resistance 
to graph de-anonymization.

\begin{figure}[!t]
\centering
\begin{subfigure}{0.49\columnwidth}
\includegraphics[width=\columnwidth]{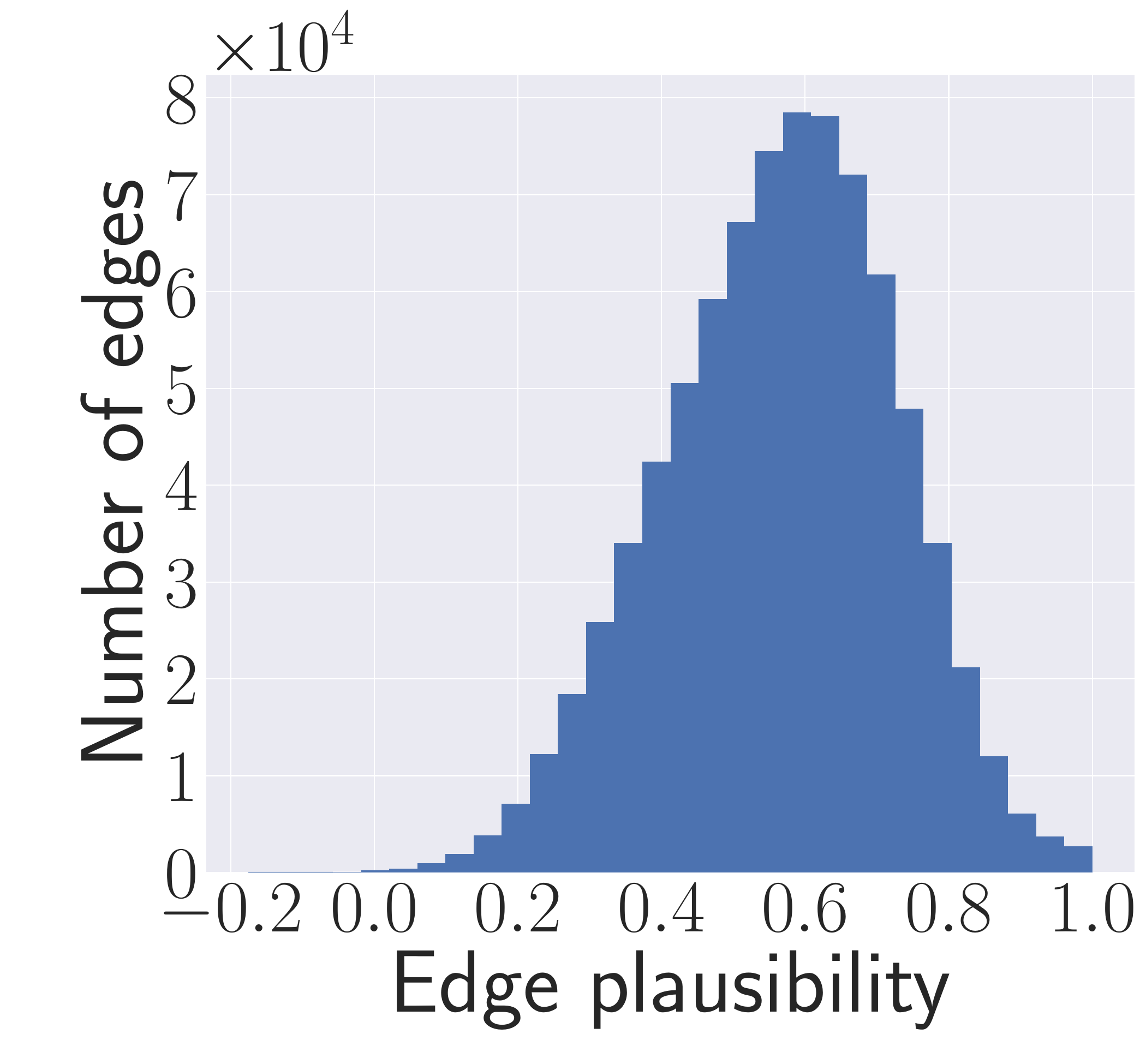}
\caption{$d=128$}
\label{fig:facebook_128_dist_hist}
\end{subfigure}
\begin{subfigure}{0.49\columnwidth}
\includegraphics[width=\columnwidth]{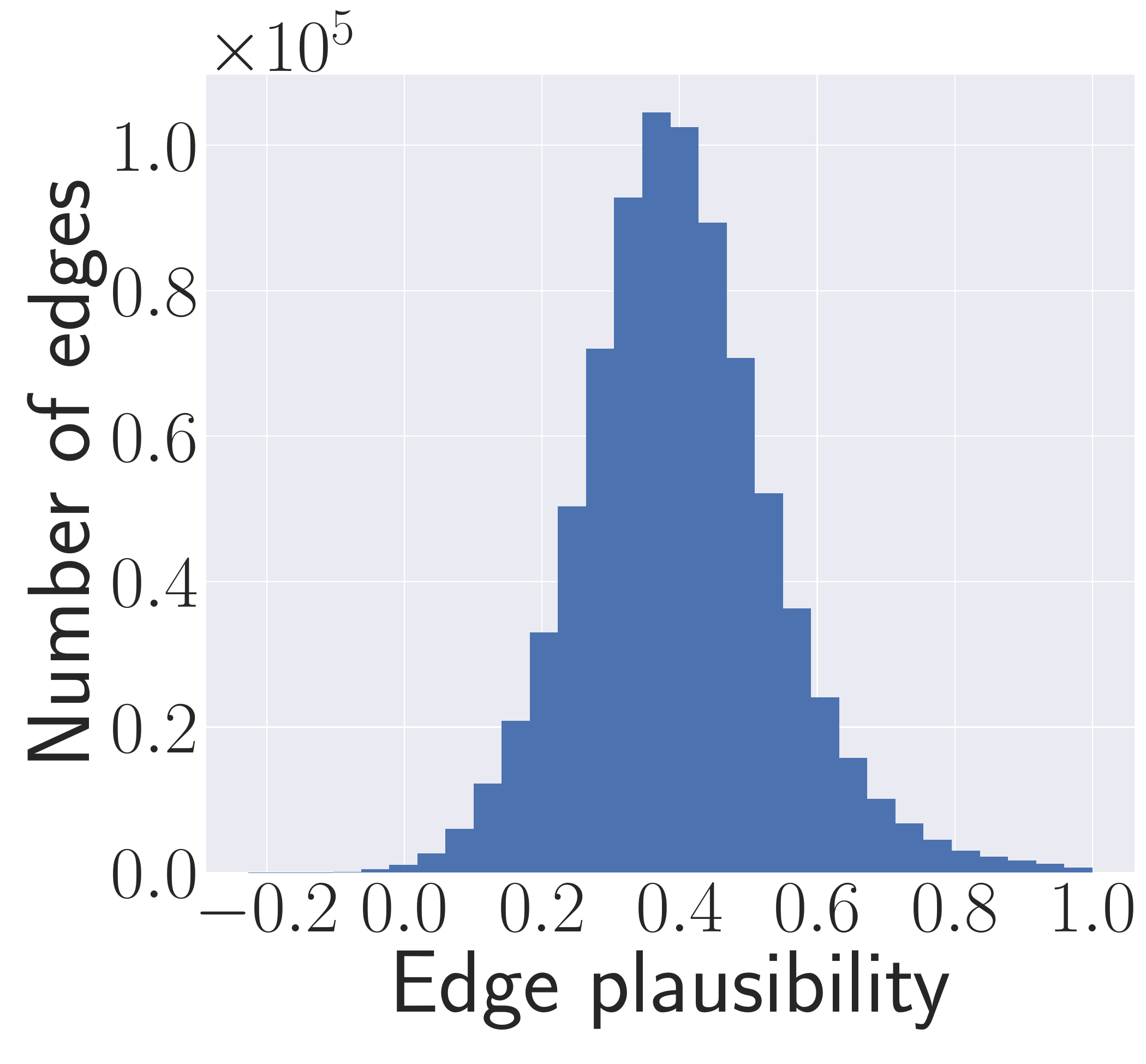}
\caption{$d=512$}
\label{fig:facebook_512_dist_hist}
\end{subfigure}
\caption{Edge plausibility
in the original NO dataset follows a Gaussian distribution.
We choose two vector dimensions for edge plausibility: (a) $d =128$ and (b) $d = 512$,
following the evaluation results in \autoref{sec:evalua}.}
\label{fig:facebook_ori_dist_hist}
\end{figure} 

\begin{figure*}[!t]
\centering
\begin{subfigure}{0.67\columnwidth}
\includegraphics[width=\columnwidth]{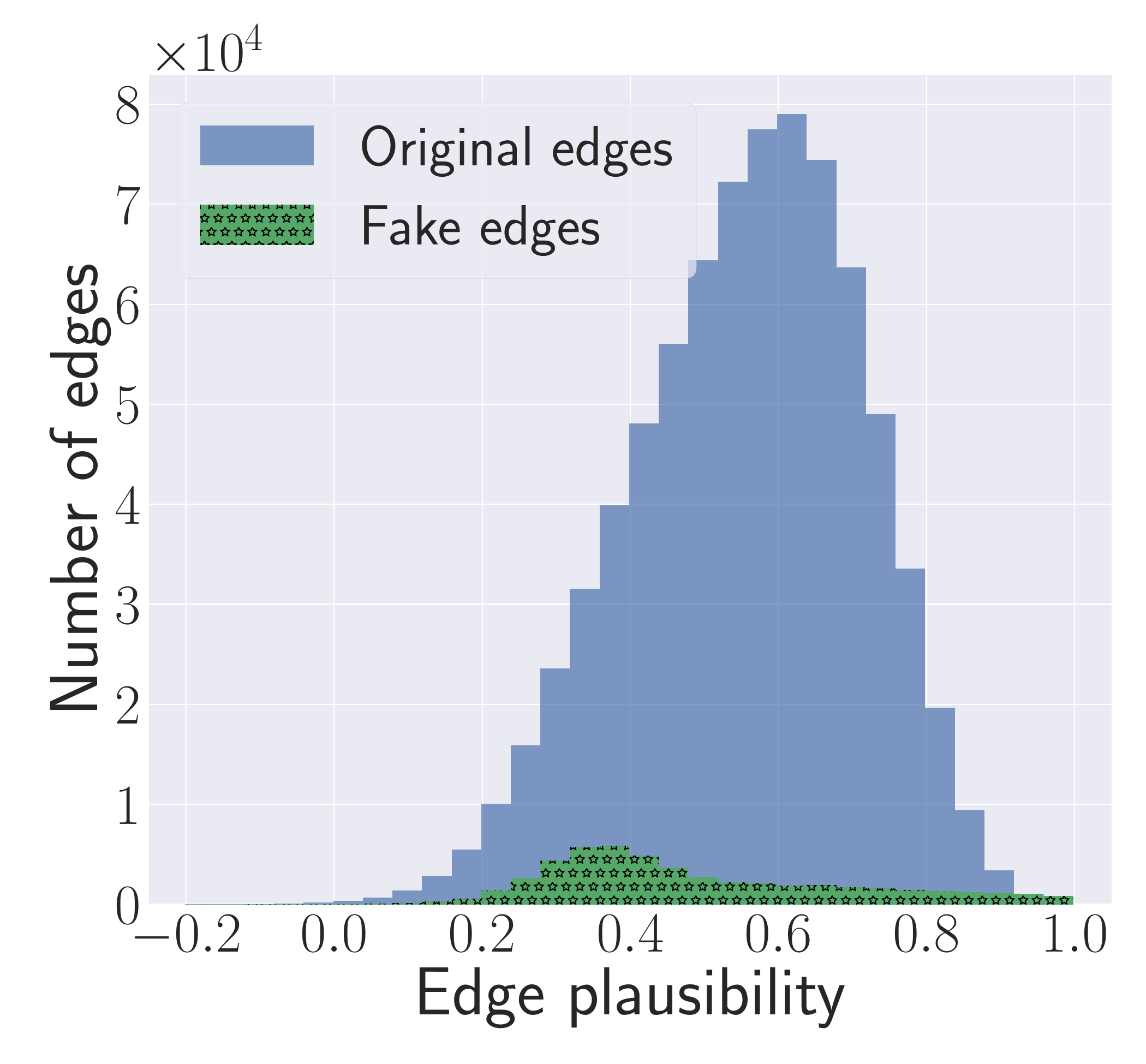}
\caption{Enhanced $k$-DA ($k=50$)}
\label{fig:facebook_ikDa_50_dist_hist}
\end{subfigure}
\begin{subfigure}{0.67\columnwidth}
\includegraphics[width=\columnwidth]{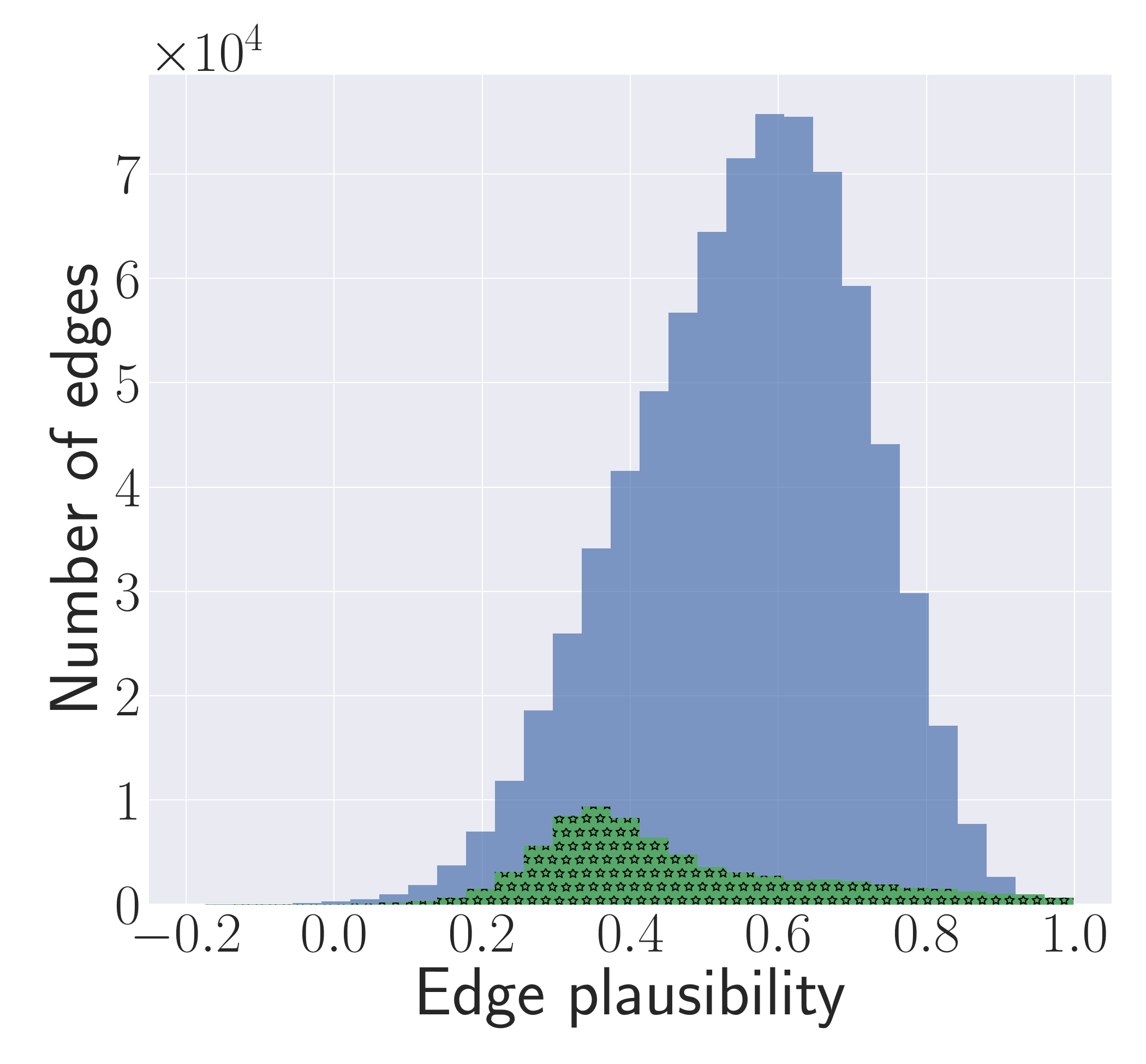}
\caption{Enhanced $k$-DA ($k=75$)}
\label{fig:facebook_ikDa_75_dist_hist}
\end{subfigure}
\\
\begin{subfigure}{0.67\columnwidth}
\includegraphics[width=\columnwidth]{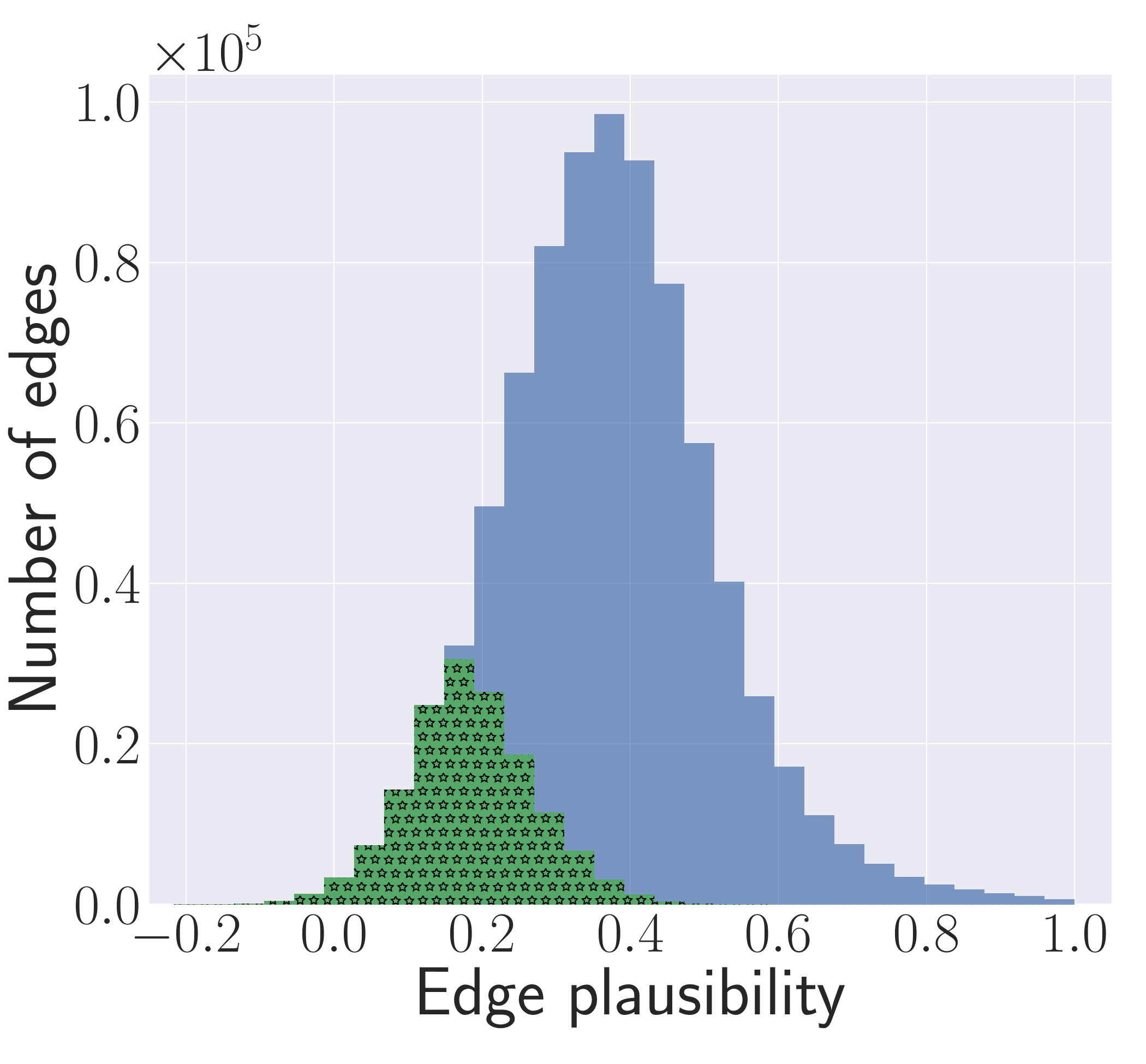}
\caption{Enhanced SalaDP ($\epsilon=100$)}
\label{fig:facebook_isdp_100_dist_hist}
\end{subfigure}
\begin{subfigure}{0.67\columnwidth}
\includegraphics[width=\columnwidth]{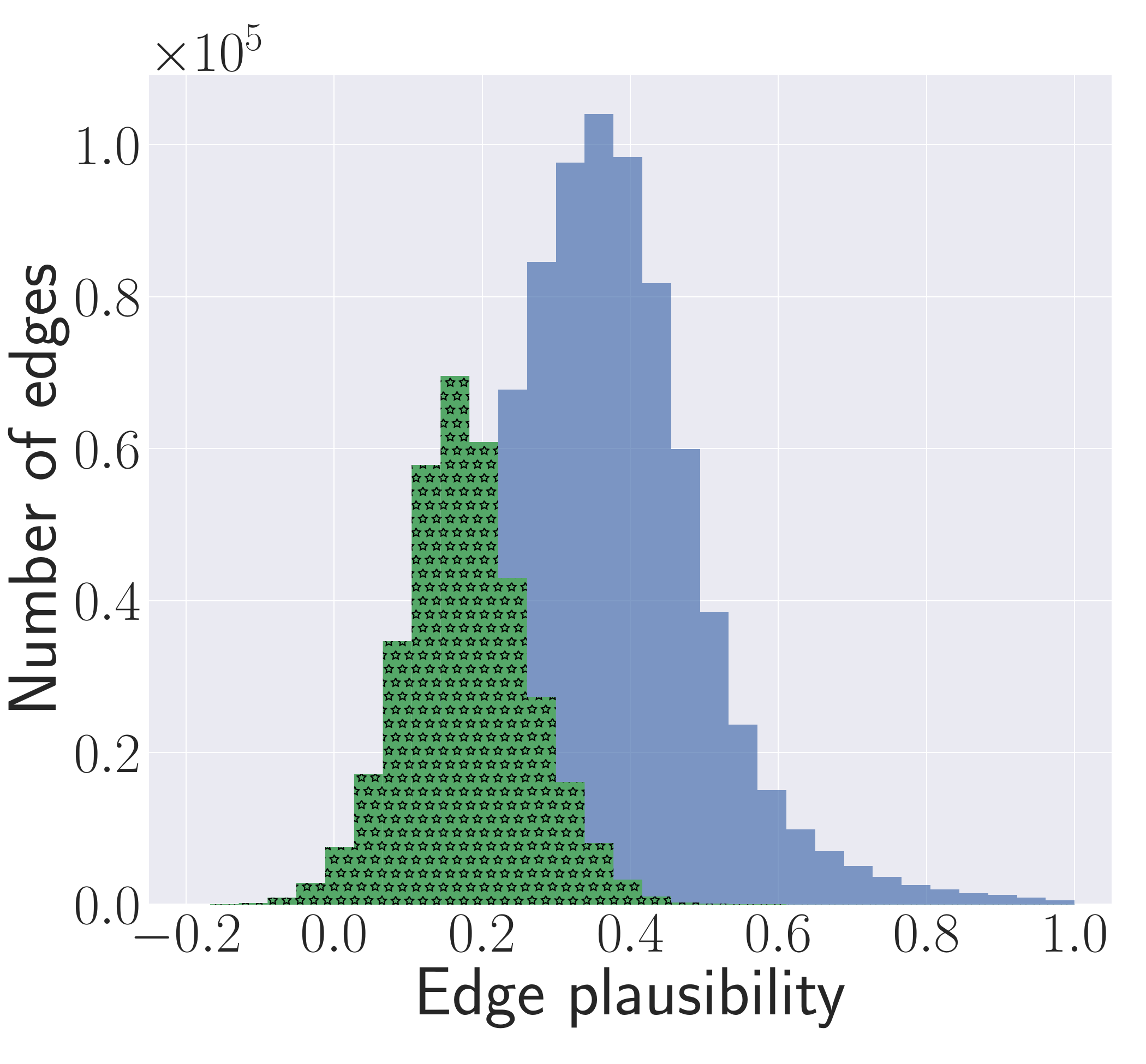}
\caption{Enhanced SalaDP ($\epsilon=50$)}
\label{fig:facebook_isdp_50_dist_hist}
\end{subfigure}
\begin{subfigure}{0.67\columnwidth}
\includegraphics[width=\columnwidth]{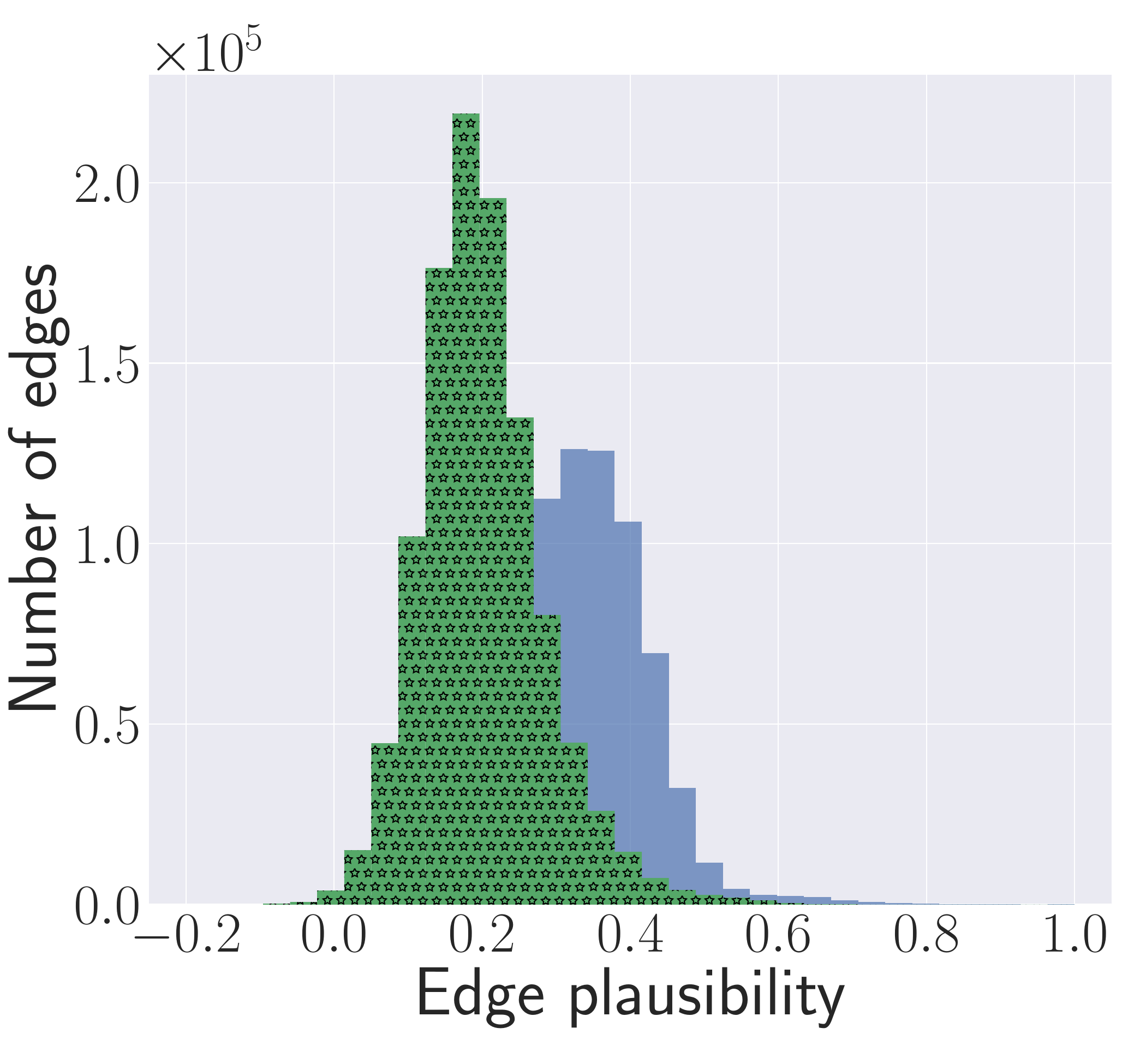}
\caption{Enhanced SalaDP ($\epsilon=10$)}
\label{fig:facebook_isdp_10_dist_hist}
\end{subfigure}
\caption{Plausibility distributions of fake and original edges 
in the NO dataset anonymized with our enhanced mechanisms.
The result for $k$-DA ($k=100$) is depicted in \autoref{fig:facebook_ikDa_100_dist_hist}.
}
\label{fig:facebook_rest_enhanced_dist_hist}
\end{figure*} 

\subsection{Methodology}

To improve the graph anonymization mechanisms,
intuitively, we should add fake edges
that are more similar to edges in the original graph $\Graph$.
\autoref{fig:facebook_ori_dist_hist}
depicts the edge plausibility distributions
for the original NO dataset under two different vector dimensions.\footnote{
We map all users in $\Graph$ into vectors
and compute all edges' plausibility in $\Graph$
following the same procedure
as for $\Graph_{\ano}$ (\autoref{sec:proxi}).}
We observe that both empirical distributions 
follow a Gaussian distribution.
If we are able to modify the current graph anonymization mechanisms
such that the plausibility of the added fake edges is more likely 
to come from the same Gaussian distribution, it should be harder to discover these fake edges.

The general procedure for 
our enhanced anonymization mechanisms is as follows.
We first apply maximum likelihood estimation 
to learn the Gaussian distribution of edge plausibility in $\Graph$,
denoted by $\mathit{N}(\proxi(\user, \user')\vert \mu,\! \sigma)$,
where $\proxi(\user, \user')$ represents $\{\user, \user'\}$'s plausibility in $\Graph$.
Then, we conduct the same process as in $k$-DA and SalaDP.
A loop is performed through all the users 
and, in each iteration, if a user $\user$ needs $m$ fake edges,
we construct a candidate set $\candi(\user)$
which includes all the potential users that could share a fake edge 
with $\user$ (following the original anonymization mechanisms' design).
Different from the original approaches of $k$-DA and SalaDP
for choosing $m$ users out of $\candi(\user)$,
we compute the plausibility between users in 
$\candi(\user)$ and $\user$,\footnote{The plausibility 
is computed over users' vectors learned from $\Graph$.} 
represented as a set $\candiproxi(\user)=\{\proxi(\user, v)\vert v\in \candi(\user)\}$.
Then, for each plausibility $\proxi(\user, v)$ in $\candiproxi(\user)$,
we calculate its density using the previously learned $\mathit{N}(\proxi(\user, \user')\vert \mu,\! \sigma)$,
and treat the density as the \emph{weight} of the user $v$ in $\candi(\user)$.
Next, we perform a weighted sampling to choose $m$ users out of $\candi(\user)$
and add edges between these users and $\user$.
In the end, we obtain our new anonymized graph $\Graph_{\fix}$ 
under the enhanced mechanisms.

Note that, as presented in \autoref{sec:preli},
for a user $\user$, SalaDP chooses $m$ users from $\candi(\user)$ in a random manner,
while $k$-DA picks the users with the highest residual degrees.
However, the reason for $k$-DA 
to take this approach is to efficiently construct the anonymized graph.
Through experiments, we discover that our enhanced $k$-DA
can also build the anonymized graph in a similar time.

We emphasize that our enhanced mechanisms 
do not affect the privacy criteria of $k$-DA and SalaDP 
as they do not modify the privacy realization process of the original mechanisms.
We will make the source code for the aforementioned enhanced versions 
of $k$-DA and SalaDP publicly available.

\subsection{Evaluation}

\mypara{Fake Edge Detection}
After obtaining $\Graph_{\fix}$,
we perform the same process as in \autoref{sec:proxi}
to compute the plausibility of all edges in $\Graph_{\fix}$.
Then, we calculate the AUC values
when using plausibility to differentiate between fake and original edges in $\Graph_{\fix}$.
The results are presented in \autoref{table:iauc}.

\begin{table}[!t]
\centering
\caption{
[Higher is better] AUC scores for detecting fake edges 
for both enhanced $k$-DA and SalaDP 
on three different datasets.
}
\label{table:iauc}
\begin{tabular}{l c c c}
\toprule
& Enron & NO & SNAP\\
\midrule
$k$-DA ($k=50$) &   0.677&0.628&0.939\\
$k$-DA ($k=75$) &   0.728&0.676&0.927\\
$k$-DA ($k=100$) & 0.753&0.702&0.896\\
\midrule
SalaDP ($\epsilon=100$) & 0.806&0.890&0.719\\
SalaDP ($\epsilon=50$) &   0.794&0.895&0.723\\
SalaDP ($\epsilon=10$) &   0.724&0.853&0.723\\
\bottomrule
\end{tabular}
\end{table}

\begin{table}[!t]
\centering
\caption{
[Higher is better] F1 scores for detecting fake edges 
using GMM and MAP estimate 
for both enhanced $k$-DA and SalaDP 
on three different datasets.
}
\label{table:if1}
\begin{tabular}{l c c c}
\toprule
& Enron & NO & SNAP\\
\midrule
$k$-DA ($k=50$) & 0.531 & 0.391 & 0.632\\
$k$-DA ($k=75$) & 0.428 & 0.433 & 0.609\\
$k$-DA ($k=100$) & 0.510 & 0.501 & 0.597\\
\midrule
SalaDP ($\epsilon=100$) & 0.422 & 0.370 & 0.515\\
SalaDP ($\epsilon=50$) & 0.390 & 0.411 & 0.522\\
SalaDP ($\epsilon=10$) & 0.439 & 0.527 & 0.490\\
\bottomrule
\end{tabular}
\end{table}

\begin{figure}[!t]
\centering
\includegraphics[width=\columnwidth]{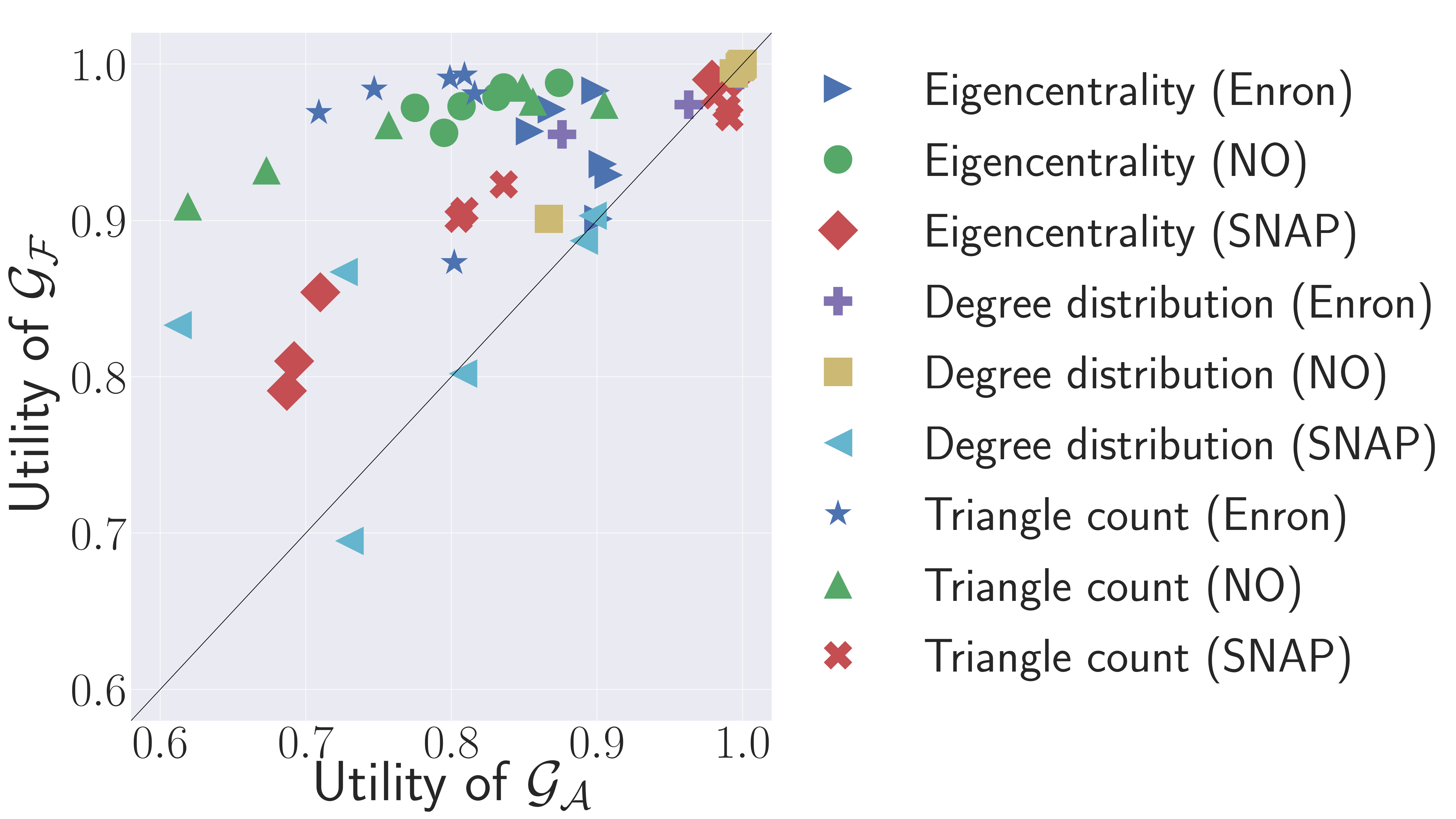}
\caption{
Comparing the utility of our enhanced mechanism ($\Graph_\fix$) 
to the original mechanism ($\Graph_\ano$) 
for different datasets and metrics.
Any point above the diagonal indicates better utility of our anonymized graph. 
The x-axis is the cosine similarity of $\Graph_\ano$ 
to the original graph $\Graph$, 
and the y-axis analogue for $\Graph_\fix$.
}
\label{fig:utility}
\end{figure}

First of all, the AUC values drop in all cases compared 
to the results in \autoref{fig:auc}.
Especially for the $k$-DA-anonymized NO dataset ($k=50$),
AUC drops by 35\% to 0.628.
This can be also observed from the histograms 
in \autoref{fig:facebook_ikDa_100_dist_hist} and \autoref{fig:facebook_rest_enhanced_dist_hist}:
By plausibility, fake edges are hidden quite well among the original edges
(compared to \autoref{fig:facebook_kDa_100_dist_hist} and \autoref{fig:facebook_rest_dist_hist}).
When applying our enhanced $k$-DA mechanism on SNAP,
the AUC values drop, but less than for NO.
This may be due to the dataset's small size (4,039 users) and the large $k$ value,
which leads to a large number of fake edges.
On the other hand,
the performance decrease for SalaDP-anonymized datasets
is smaller, but still significant.

Moreover, we discover from \autoref{fig:facebook_ikDa_100_dist_hist} 
and \autoref{fig:facebook_rest_enhanced_dist_hist} that
the two Gaussian distributions of $\Graph_{\fix}$ 
for $k$-DA and SalaDP largely overlap 
(see \autoref{fig:facebook_kDa_100_dist_hist} and \autoref{fig:facebook_rest_dist_hist} for comparison).
This indicates that the Gaussian mixture model approach described in \autoref{sec:evalua} 
cannot perform effective fake edge detection.
For instance, our experiments with the GMM approach only achieve around 0.37 F1 score
for SalaDP ($\epsilon=100$) on the NO dataset, 
which represents a 50\% performance drop (see~\autoref{table:if1}).

It is worth noting that all the edges added by our enhanced anonymization mechanisms
still have relatively smaller plausibility than the original edges.
Given that our weighted sampling 
follows the original edges' plausibility distribution in $\Graph$,
this implies that not many potential fake edges are normal
with respect to plausibility.
We conclude that it is non-trivial to create fake edges totally indistinguishable from original edges.

\mypara{Graph Utility}
The main motivation for OSNs
to share their graph data is to allow third parties
to conduct research or build commercial applications.
Therefore, a graph anonymization mechanism
needs to take into account graph utility,
i.e., how well the anonymized graph preserves the structural properties of the original graph.
To show that our enhanced mechanisms
outperform the current anonymization mechanisms,
we also evaluate $\Graph_{\fix}$'s utility.

There exist many graph properties that 
can be used to evaluate graph utility~\cite{EK10,LRU14,JLMHB15}. 
For the sake of conciseness, 
we focus on three of them including degree distribution, eigencentrality, and triangle count.
The degree distribution represents the proportion of users
with a certain degree for all possible degrees.
Eigencentrality evaluates the influence/importance of each user in a graph. 
It assigns a centrality score for each user 
based on the eigenvector of the graph's adjacency matrix.
Triangle count summarizes the number of triangles each user belongs to in a graph
which reflects the graph connectivity~\cite{LRU14}.
We compute the three properties for $\Graph$, $\Graph_{\ano}$, and $\Graph_{\fix}$, 
and calculate the cosine similarity 
between $\Graph$'s and $\Graph_{\ano}$'s properties 
as well as between $\Graph$'s and $\Graph_{\fix}$'s properties.
Higher similarity naturally implies better utility.

\autoref{fig:utility} presents the results.
We first observe a strong similarity between $\Graph_{\fix}$ and $\Graph$
for all graph properties, i.e., $\Graph_{\fix}$ preserves high utility.
For instance, the cosine similarity for triangle count is above 0.86
in most of the cases.
Meanwhile, the lowest cosine similarity (degree distribution)
is still approaching 0.7 when applying enhanced $k$-DA ($k=100$)
to SNAP.

More importantly, we observe that 
$\Graph_{\fix}$ preserves better graph utility than $\Graph_{\ano}$
(almost all points in \autoref{fig:utility} are above the diagonal).
For instance, the eigencentrality's cosine similarity 
between $\Graph_{\fix}$ and $\Graph$
is 0.985 while the similarity between $\Graph_{\ano}$ and $\Graph$ is only 0.836
for the $k$-DA-anonymized NO dataset ($k=50$).
This is because the fake edges added by our enhanced mechanisms
are more structurally similar to the original edges, thus preserving better utility.

\mypara{Graph De-anonymization}
Next, we investigate the performance of graph de-anonymization
on graphs generated by our enhanced mechanisms.
We concentrate on the NS-attack~\cite{NS09} 
due to its superior performance over others~\cite{JLMHB15}.
The NS-attack assumes that the adversary knows an \emph{auxiliary graph}
with all nodes' identities.
Her goal is to map each node in the auxiliary graph
to the node representing the same user in an anonymized \emph{target graph}.
Correctly matched nodes are thus successfully de-anonymized in the target graph.
To ease this matching, the NS-attack assumes that the adversary has prior knowledge
of some correctly matched nodes, namely the \emph{seed nodes}. The attack then 
starts from these seeds to de-anonymize more nodes
by propagating throughout 
the whole anonymized graph.

We use $\Graph_{\ano}$ and $\Graph_{\fix}$ as the target graphs, respectively,
and sample a subgraph from the original graph $\Graph$ 
containing all edges among 25\% randomly selected nodes in $\Graph$
as the auxiliary graph.
Moreover, we choose the 200 nodes with the highest degrees 
from the auxiliary graph as our seeds.\footnote{We tried other sampling approaches for seed nodes,
but did not observe significant performance differences.}
For evaluation, 
we concentrate on correctly and wrongly de-anonymized users.

\begin{table}[!t]
\centering
\caption{
De-anonymization prevention of our enhanced mechanism ($\Graph_{\fix}$) 
and the original mechanism ($\Graph_{\ano}$). 
[Lower is better] Number of nodes the NS-attack can correctly  
de-anonymize. Best scores are in bold.
}
\label{table:deanony}
\setlength{\tabcolsep}{5pt} 
\begin{tabular}{l c c c c c c}
\toprule
& \multicolumn{2}{c}{Enron} & \multicolumn{2}{c}{NO} & \multicolumn{2}{c}{SNAP}\\
\midrule
& 
$\Graph_{\ano}$ &
$\Graph_{\fix}$ &
$\Graph_{\ano}$ &
$\Graph_{\fix}$ &
$\Graph_{\ano}$ &
$\Graph_{\fix}$ \\
\midrule
$k$-DA ($k=50$) & 307 & 289 & 759 & \textbf{532} & 328 & 303\\
$k$-DA ($k=75$) & 309 & 270 & 689 & 508 & 294 &234\\
$k$-DA ($k=100$) & 302 & \textbf{256} & 580 & 491 & 274 & \textbf{208}\\
\midrule
SalaDP ($\epsilon=100$) & 265 & 255 & 470 & \textbf{396} & 378 & 342\\
SalaDP ($\epsilon=50$) & 243 & 225 & 291 & 277 & 370 & 290\\
SalaDP ($\epsilon=10$) & 236 & \textbf{207} & 233 & 208 & 376 & \textbf{267}\\
\bottomrule
\end{tabular}
\end{table}

\begin{figure}[!t]
\centering
\includegraphics[width=\columnwidth]{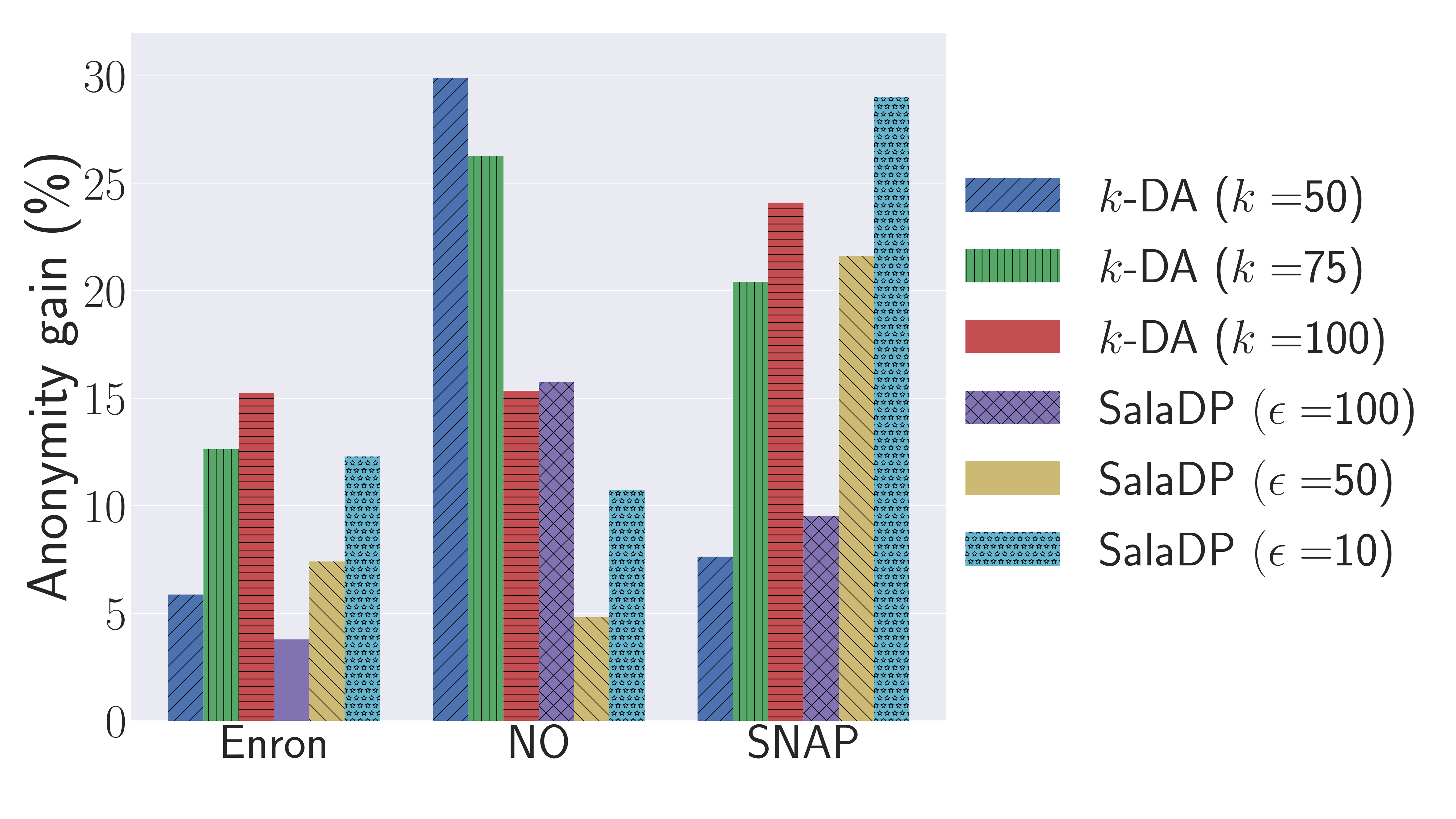}
\caption{
[Higher is better] Gain in anonymity with respect to the reduction of the number of 
correctly de-anonymized nodes by the NS-attack
between using our enhanced mechanism versus 
the original mechanism.
}
\label{fig:deanony_drop}
\end{figure}

\autoref{table:deanony} shows the results.
First of all, the number of correctly de-anonymized nodes
by the NS-attack is reduced in all cases thanks to our enhanced mechanisms.
\autoref{fig:deanony_drop} 
further depicts the anonymity gain,
i.e., the performance drop with respect to the correctly de-anonymized nodes.
We see that the NS-attack de-anonymizes almost 30\% fewer nodes
on the enhanced $k$-DA-anonymized ($k$=50) NO dataset.
We also notice from \autoref{table:deanony} that our enhanced mechanisms
reduce the total number of nodes 
that the NS-attack de-anonymizes (both correct and wrong ones).
This indicates that the NS-attack's ability to propagate 
also degrades in graphs anonymized 
by our enhanced mechanisms. 
\section{Related Work}
\label{sec:relwork}

Various graph anonymization mechanisms have been 
proposed in the literature~\cite{LT08,ZP08,HMJTW08,BCKS09,ZCO09,TY09,
CFL10,PGM12,WW13,XCT14,MPS13}.
One class of these mechanisms 
follows the concept of $k$-anonymity.
Liu and Terzi~\cite{LT08} propose the first approach in this direction,
i.e., $k$-DA, which we concentrate on in this paper. 
Meanwhile, Zhou and Pei~\cite{ZP08} propose $k$-neighborhood anonymity,
where each user in the anonymized graph shares the same neighborhood,
i.e., the sub-social network among her friends,
with at least $k-1$ other users.
The authors adopt minimum BFS coding to represent each user's neighborhood,
then rely on a greedy match to realize $k$-neighborhood anonymity.

Another class of graph anonymization mechanisms 
is inspired by differential privacy.
Besides SalaDP,
multiple solutions have been proposed~\cite{PGM12,WW13,XCT14}.
For instance, Wang and Wu~\cite{WW13} present 
a 2K-graph generation model to achieve differential privacy,
where noise is added based on smooth sensitivity.
Xiao et al.~\cite{XCT14} encode users' connection probabilities
with a hierarchical random graph model,
and perform Markov chain Monte Carlo 
to sample a possible graph structure from the model
while enforcing differential privacy.
Besides the above,
other graph anonymization techniques include~\cite{HMJTW08,BCKS09,MPS13}.

Due to space constraints, 
we only consider the two most widely known anonymization mechanisms,
i.e., $k$-DA and SalaDP.
In the future, we plan to apply our approach
to more anonymization mechanisms.

Besides anonymization, graph de-anonymization 
has been extensively studied as well.
Backstrom et al.
are among the first to de-anonymize users in a naively anonymized social graph~\cite{BDK07}.
The attack of Narayanan and Shmatikov
is essentially a framework~\cite{NS09},
based on which multiple approaches 
have been proposed~\cite{SH12,JLSB14,NKA14,SD14}.
We emphasize that graph de-anonymization is orthogonal to our graph recovery attack.
First of all, graph de-anonymization attacks aim to identify users in an anonymized graph
while our graph recovery aims to find fake added edges.
As shown in~\autoref{sec:prideg},
our graph recovery can degrade anonymized graphs' privacy guarantees.
The reason our graph recovery cannot increase the performance of graph de-anonymization (in our case, of the NS-attack),
is that most of the graph de-anonymization attacks assume a much stronger attack model
than those considered in graph anonymization mechanisms.
Therefore, we propose privacy loss metrics tailored to $k$-DA and SalaDP, which we believe are more appropriate. 
Moreover, we show that our enhanced anonymization mechanisms 
that are inspired by our graph recovery attack significantly reduce the success rate of graph de-anonymization.
\section{Conclusion}
\label{sec:conclu}

In this paper,
we identify a fundamental vulnerability of 
the existing graph anonymization mechanisms which do not take into account 
key structural characteristics of a social graph
when adding fake edges to it.
We propose an edge plausibility metric based on graph embedding 
that enables us to exploit this weakness in order to identify fake edges.
Extensive experiments show that,
using this metric, we are able to recover the original graph 
from an anonymized graph to a large extent.
Our graph recovery also results in significant 
privacy damage to the original anonymization mechanisms.
To mitigate this weakness, 
we propose enhancement over the existing anonymization mechanisms.
Our experiments show that our enhanced mechanisms 
significantly reduce the performance of our graph recovery attack,
increase graph de-anonymization resistance,
and at the same time provide better graph utility.

\section*{Acknowledgment}
We thank the anonymous reviewers, 
and our shepherd, Anupam Das, for their helpful feedback and guidance.

\balance
\bibliographystyle{IEEEtranS}
\bibliography{normal_generated}

\end{document}